\documentclass[aps,11pt,onecolumn,preprintnumbers,showpacs,showkeys,nofootinbib,
superscriptaddress,prd]{revtex4-1}
\usepackage[T1]{fontenc}
\usepackage[american]{babel}
\usepackage{epsfig}
\usepackage{amssymb,amsmath,amsfonts,amsthm,graphicx,psfrag}
\usepackage{hyperref}

\newcommand{\beq}{\begin{eqnarray}}
\newcommand{\eeq}{\end{eqnarray}}

\newcommand{\tr}{\operatorname{Tr}}

\begin{document}
\preprint{}

\title{
Lattice study of static quark-antiquark interactions in dense quark matter
}

\author{N.\,Yu.~Astrakhantsev}
\email[]{nikita.astrakhantsev@itep.ru}
\affiliation{ Moscow Institute of Physics and Technology, Dolgoprudny, 141700 Russia }
\affiliation{ Institute for Theoretical and Experimental Physics NRC ``Kurchatov Institute'', Moscow, 117218 Russia }

\author{V.\,G.~Bornyakov}
\email[]{vitaly.bornyakov@ihep.ru}
\affiliation{ NRC ``Kurchatov Institute'' - IHEP, 142281 Protvino, Russia }
\affiliation{ Far Eastern Federal University, School of Biomedicine, 690950 Vladivostok, Russia }

\author{V.\,V.~Braguta}
\email[]{braguta@itep.ru}
\affiliation{ Moscow Institute of Physics and Technology, Dolgoprudny, 141700 Russia }
\affiliation{ Institute for Theoretical and Experimental Physics NRC ``Kurchatov Institute'', Moscow, 117218 Russia } 
\affiliation{ Far Eastern Federal University, School of Biomedicine, 690950 Vladivostok, Russia }
\affiliation{ Bogoliubov Laboratory of Theoretical Physics, Joint Institute for Nuclear Research, Dubna, 141980 Russia } 

\author{E.-M.~Ilgenfritz}
\affiliation{ Bogoliubov Laboratory of Theoretical Physics, Joint Institute for Nuclear Research, Dubna, 141980 Russia }

\author{A.\,Yu.~Kotov}
\email[]{kotov@itep.ru}
\affiliation{ Moscow Institute of Physics and Technology, Dolgoprudny, 141700 Russia }
\affiliation{ Institute for Theoretical and Experimental Physics NRC ``Kurchatov Institute'', Moscow, 117218 Russia }
\affiliation{ Bogoliubov Laboratory of Theoretical Physics, Joint Institute for Nuclear Research, Dubna, 141980 Russia }

\author{A.\,A.~Nikolaev}
\email[]{Alexander.Nikolaev@itep.ru}
\affiliation{ Institute for Theoretical and Experimental Physics NRC ``Kurchatov Institute'', Moscow, 117218 Russia }

\author{A.~Rothkopf}
\email[]{alexander.rothkopf@uis.no}
\affiliation{Institut f\"ur Theoretische Physik, Universit\"at Heidelberg, Philosophenweg 12, 69120 Heidelberg, Germany}
\affiliation{Faculty of Science and Technology, University of Stavanger, NO-4036 Stavanger, Norway}

\begin{abstract}
In this paper we study the interactions among a static quark-antiquark pair in the presence of dense two-color quark matter with lattice simulation.
To this end we compute Polyakov line correlation functions and determine the renormalized color averaged, color singlet and color triplet grand potentials. The color singlet grand potential allows us to elucidate the number of quarks induced by a static quark antiquark source, as well as the internal energy of such a pair in dense quark matter. We furthermore determine the screening length, which in the confinement phase is synonymous with the string breaking distance. The screening length is a decreasing function of baryon density, due to the possibility to break the interquark string via a scalar diquark condensate at high density. 
We also study the large distance properties of the color singlet grand potential in a dense medium and find that it is well described by a simple Debye screening formula, parameterized by a Debye mass and an effective coupling constant. The latter is of order of unity, i.e. even at large density two-color quark matter is a strongly correlated system.
\end{abstract}



\maketitle

\section{Introduction}

Knowledge of the properties of QCD at large baryon density is needed to interpret the results of heavy ion collisions experiments. 
In particular, this is the case at the future experiments of NICA (JINR, Dubna) and FAIR (Darmstadt, Germany), which are designed to study the region of high baryon density. Input from the theory side is hence urgently needed. An understanding of the properties of matter in the corresponding region of the QCD phase diagram is also extremely important in astrophysics, for example, for a correct description of the fusion of neutron stars.

In general, lattice QCD is a powerful tool to study the nonperturbative properties of strongly interacting matter from first principles. 
By virtue of lattice simulations vital insight into QCD at finite temperature~\cite{Borsanyi:2016bzg}, nonzero magnetic field~\cite{DElia:2013xrj},
isospin chemical potential~\cite{Brandt:2017oyy, Scior:2017cne} and chiral chemical potential~\cite{Braguta:2015zta, Braguta:2015owi, Braguta:2016aov} has been obtained.

An interesting area of finite temperature lattice simulations is the study of 
the interaction between a quark-antiquark pair and the interaction of the 
pair with the QCD medium (see e.g.~\cite{Kaczmarek:2002mc, Kaczmarek:2004gv, Kaczmarek:2005ui, Kaczmarek:2005gi,Maezawa:2010vj,Maezawa:2011aa}). 

The in-medium properties of QCD are prominently encoded in the correlation function of Polyakov loops (i.e. Polyakov lines with maximum temporal extent). As the Polyakov loop constitutes a quasi order parameter of the strong interactions, its correlator is greatly affected by the phase transitions which take place in QCD. One pertinent example is the confinement/deconfinement phase 
transition which considerably modifies the value of the correlator.

Furthermore the Polyakov loop correlator is directly related to the free energy of the in-medium quark-antiquark pair.In the confinement phase the free energy extracted is known to be a linear increasing function at intermediate distances. At zero temperature and distances $\sim 1.2~$fm the free energy asymptotes to a plateau due to the string breaking phenomenon. On the other hand in the deconfinement phase at large distances the free energy also flattens off, the reason being a screening of the interactions between the quark and antiquark due to liberated colored medium degrees of freedom. The question of whether or how the screening properties of QCD may be captured by an analogous and equally simple Debye screening formula in analogy with the Abelian theory is a ongoing field of research. The properties of the correlation function of Polyakov loops at finite temperature in QCD have been thoroughly studied in lattice simulations~\cite{Kaczmarek:2002mc, Kaczmarek:2004gv, Kaczmarek:2005ui, Kaczmarek:2005gi,Maezawa:2010vj,Maezawa:2011aa}. More recently the Polyakov loop correlator on the lattice has been compared to effective field theory predictions, both in a perturbative setting in pNRQCD, as well as perturbatively matched EQCD~\cite{Bazavov:2018wmo}. For analytic studies of the Polyakov loop see e.g.~\cite{Fischer:2013eca,Agasian:2017tag}. 

While it is an interesting proposition to carry out similar studies of the Polyakov loop at finite Baryon density in QCD, the usual methods of lattice QCD unfortunately break down because of the so-called sign problem. Instead approaches, such as the Taylor expansion or analytical continuation (both in quark chemical potential) allow one to obtain useful results at small values of the chemical potential (see e.g.~\cite{Bazavov:2017dus,Gunther:2016vcp,DElia:2016jqh} ). 
There are a lot of analytical attempts to study properties of dense matter (see e.g.~\cite{Buballa:2003qv, Alford:2007xm, Andreichikov:2017ncy}. However, most of them are not based on first principles and it is not clear how to reliably estimate the systematic uncertainty of different models.

Instead of pursuing the question of finite density physics in QCD directly, we here turn to the study of theories, which are similar to QCD but are not plagued by the sign problem. We believe that in particular the study of dense two-color QCD~\cite{Kogut:1999iv,Kogut:2000ek} allows us to learn about the properties of regular QCD at nonzero chemical potential. Other candidate theories not further pursued here are e.g. QCD at nonzero isospin chemical potential~\cite{Son:2000xc, Janssen:2015lda}. Of course we cannot expect to obtain quantitative predictions from such a strategy, while vital qualitative insight may be gained.

Two-color QCD at finite chemical potential has been studied with lattice simulations quite intensively before, see, e.g.~\cite{Hands:1999md,Muroya:2002ry,Kogut:2002cm, Cotter:2012mb, Braguta:2016cpw, Holicki:2017psk} and references therein. 
Mostly these papers are aiming at the study of the phase diagram of two-color QCD in the region of small and moderate baryon densities. 

The phase structure of two-color QCD at large baryon densities was studied in our previous paper~\cite{Bornyakov:2017txe}, where lattice simulations were carried out at a relatively small lattice spacing $a=0.044$ fm. Compared to previous works, it allows us to extent the range of accessible values of the baryon density, up to quark chemical potential $\mu>2000~$MeV, avoiding strong lattice artifacts. The main result of the paper~\cite{Bornyakov:2017txe} is the observation of the confinement/deconfinement transition at finite density and low temperature.  
In view of this finding we are interested in studying the properties of the Polyakov loop correlation function in cold dense quark matter and to shed light on how they are affected by the confinement/deconfinement transition, which takes place at finite density. This is the central question addressed in this paper.

The manuscript is organized as follows. In the next section~\ref{sec:simdet} we describe our lattice set-up. In section~\ref{sec:phasediag} we give an overview of the present status of the cold dense two-color QCD phase diagram. 
Section~\ref{sec:grandpot} is devoted to the calculation of the renormalized Polyakov loop correlation functions, as well as of the
color averaged, color singlet and color triplet grand potentials. In addition, in the same section we determine the renormalized Polyakov loop and the grand potential of a single quark/antiquark.
Using the color singlet grand potential, the quark number and internal energy induced by a static quark-antiquark pair are obtained in section~\ref{sec:quarknum}. We consider the string breaking phenomenon in dense quark matter in section~\ref{sec:stringbreak} and Debye screening in section~\ref{sec:DebyeScreen}. In the last section~\ref{sec:conslusion} we discuss our results and conclude.

\section{Simulation details}
\label{sec:simdet}

In our lattice study we used the tree level improved Symanzik gauge action~\cite{Weisz:1982zw, Curci:1983an}. For the fermionic degrees of freedom we used staggered fermions with an action of the form
\begin{equation}
\label{eq:S_F}
S_F = \sum_{x, y} \bar \psi_x M(\mu, m)_{x, y} \psi_y + \frac{\lambda}{2} \sum_{x} \left( \psi_x^T \tau_2 \psi_x + \bar \psi_x \tau_2 \bar \psi_x^T \right)
\end{equation}
with
\begin{equation}
\label{eq:Dirac_operator}
M(\mu,m)_{xy} = ma\delta_{xy} + \frac{1}{2}\sum_{\nu = 1}^4 \eta_{\nu}(x)\Bigl[ U_{x, \nu}\delta_{x + h_{\nu}, y}e^{\mu a\delta_{\nu, 4}} - U^\dagger_{x - h_{\nu}, \nu}\delta_{x - h_{\nu}, y}e^{- \mu a\delta_{\nu, 4}} \Bigr]\,,
\end{equation}
where $\bar \psi$, $\psi$ are staggered fermion fields, $a$ is the lattice spacing, $m$ is the bare quark mass, and $\eta_{\nu}(x)$ are the standard staggered phase factors: $\eta_1(x) = 1,\, \eta_\nu(x) = (-1)^{x_1 + ...+ x_{\nu-1}},~\nu=2,3,4$.
The chemical potential $\mu$ is introduced into the Dirac operator ~\eqref{eq:Dirac_operator} through the multiplication of the links along
and opposite to the temporal direction by factors $e^{\pm \mu a}$ respectively. 
This way of introducing
the chemical potential makes it possible to avoid additional divergences and to reproduce well known
continuum results~\cite{Hasenfratz:1983ba}.

In addition to the standard staggered fermion action we add 
a diquark source term~\cite{Hands:1999md}
to equation (\ref{eq:S_F}). The diquark source term explicitly violates $U_V(1)$ and allows to observe diquark 
condensation even on finite lattices, because this term effectively chooses one vacuum from the family of $U_V(1)$-symmetric vacua. 
Typically one carries out numerical simulations at a few nonzero values of the parameter $\lambda$ and then extrapolates to zero $\lambda$. 
Notice, however, that this paper is aimed at studying the region of large baryon density where lattice simulations are numerically very expensive. For this reason, in this paper we have chosen a different strategy. Most of our lattice simulations are conducted at a single fixed value $\lambda=0.00075$. 
In order to check 
the $\lambda$-dependence of our results for chemical potentials $a\mu=0.0,\,0.1,\,0.2,\,0.3,\,0.4$ we carry out additional lattice simulations 
for values of  $\lambda=0.0005,\,0.001$.  

Integrating out the fermion fields, the partition function for the theory with the action $S=S_G+S_F$
can be written in the form
\beq
Z = \int DU e^{-S_G} \cdot Pf \begin{pmatrix} \lambda \tau_2 & M \\ -M^T & \lambda \tau_2 \end{pmatrix} = \int DU e^{-S_G} \cdot {\bigl ( \det (M^\dagger M + \lambda^2) \bigr )}^{\frac 1 2},
\label{z1}
\eeq
which corresponds to $N_f=4$ dynamical fermions in the continuum limit. Note that the pfaffian $Pf$ is strictly positive, such that one can use 
Hybrid Monte-Carlo methods to study this system.
First lattice studies of the theory with partition function (\ref{z1}) have been carried 
out in the following papers~\cite{Kogut:2001na, Kogut:2001if, Kogut:2002cm}. In 
the present study we are going to investigate instead a theory with the partition function
\beq
Z=\int DU e^{-S_G} \cdot {\bigl ( \det (M^\dagger M + \lambda^2) \bigr )}^{\frac 1 4},
\label{z2}
\eeq
which corresponds to $N_f=2$ dynamical fermions in the continuum limit.

It is known that the symmetries of the staggered fermion action are different from those of two-color QCD
with fundamental quarks~\cite{Hands:1999md}. In particular, the symmetry breaking pattern of QC$_2$D
with fundamental quarks is SU(2$N_f$) $\to$ Sp(2$N_f$), whereas for staggered quarks it is SU(2$N_f$) $\to$ O(2$N_f$). 
Notice, however, that in the naive continuum limit for the staggered action with the diquark source term, the action factorizes into two copies of $N_f=2$ fundamental fermions\cite{Braguta:2016cpw}.
In addition, for sufficiently small lattice spacing $a$ the $\beta$-function of the theory (\ref{z2}) measured in \cite{Braguta:2016cpw} corresponds
to the $\beta$-function of QC$_2$D with two fundamental flavors. For these reasons we believe, that
the partition function (\ref{z2}) after the rooting procedure corresponds to QC$_2$D with $N_f = 2$ fundamental fermions with a correct 
continuum chiral symmetry breaking pattern.

The results presented in this paper have been obtained in lattice simulations performed on a $32^4$ lattice for a set of the chemical potentials in the region $a\mu \in (0, 0.5)$.
At zero density we performed scale setting using the QCD Sommer scale $r_0=0.468(4) \mathrm{~fm}$ ~\cite{Bazavov:2011nk}.
In this case the string tension associated to $\mu_q=0$ amounts to $\sqrt{\sigma_0}=476(5) \mathrm{~MeV}$ at $a = 0.044 \mathrm{~fm}$. 

Numerical simulations in the region of large baryon density require 
considerable computer resources. For this reason, for the present paper 
we performed our study at a pion mass of $m_{\pi}=740(40) \mathrm{~MeV}$, where the cost is manageable. We will preferentially choose a smaller pion mass in future simulations.

To calculate Wilson loops we have employed the following smearing scheme: one step of HYP smearing~\cite{Hasenfratz:2001hp} for temporal links with the smearing parameters according to the HYP2 parameter set~\cite{DellaMorte:2005nwx} followed by 100 steps of APE smearing~\cite{Albanese:1987ds} (for spatial links only) with a smearing parameter $\alpha_{APE} = 2/3$. We found that the HYP2 parameter set provides a better signal-to-noise ratio for Wilson loops and correlators of the Polyakov loops than the HYP1 set. As for the calculations of Polyakov loop, color-averaged~(\ref{eq:Omega_av}), color-singlet~(\ref{eq:Omega_1}) and color-triplet~(\ref{eq:Omega_3}) correlators one step of HYP2 smearing for temporal links was performed, but in the case of singlet and triplet correlators the Coulomb gauge without residual gauge fixing was fixed at first. The formulas used in the calculation are 
\begin{eqnarray}
\label{eq:Omega_av}
\text{exp}\left[- \frac{\Omega_{\bar q q}(r, \mu)}{T}\right] &=& \frac{1}{4} \left\langle \text{Tr} L(\vec{r}) \text{Tr} L^\dagger(0) \right\rangle\,, \\
\label{eq:Omega_1}
\text{exp}\left[- \frac{\Omega_1(r, \mu)}{T}\right] &=& \frac{1}{2} \left\langle \text{Tr} L(\vec{r}) L^\dagger(0) \right\rangle\,, \\
\label{eq:Omega_3}
\text{exp}\left[- \frac{\Omega_3(r, \mu)}{T}\right] &=& \frac{1}{3} \left\langle \text{Tr} L(\vec{r}) \text{Tr} L^\dagger(0) \right\rangle - \frac{1}{6} \left\langle \text{Tr} L(\vec{r}) L^\dagger(0) \right\rangle\,.
\end{eqnarray}
The reason why we performed smearing is that, due to the large time extension ($L_t = 32$) correlators of the Polyakov loops by default are very noisy. Thus one has to introduce some smoothing technique to extract the signal. An analogous smearing scheme was applied in the papers~\cite{Bonati:2017uvz,Bonati:2014ksa}. 

\section{The phase diagram of dense two-color QCD at low temperatures}
\label{sec:phasediag}

Let us explore the tentative phase structure of two-color QCD as basis for our study of interquark interactions. Based on symmetry arguments it is possible to build a chiral perturbation theory (ChPT) for sufficiently small chemical potential~\cite{Kogut:1999iv, Son:2000xc, Kogut:2000ek}.
This ChPT can be used to predict the phase transitions at sufficiently small values of 
the chemical potential. In particular, it was predicted that for small
values of chemical potential ($\mu<m_{\pi}/2$) the system is in the hadronic phase. 
In this phase the system exhibits confinement and chiral symmetry is broken. 

\begin{figure}[hb]
\centering
	\includegraphics[scale=0.8]{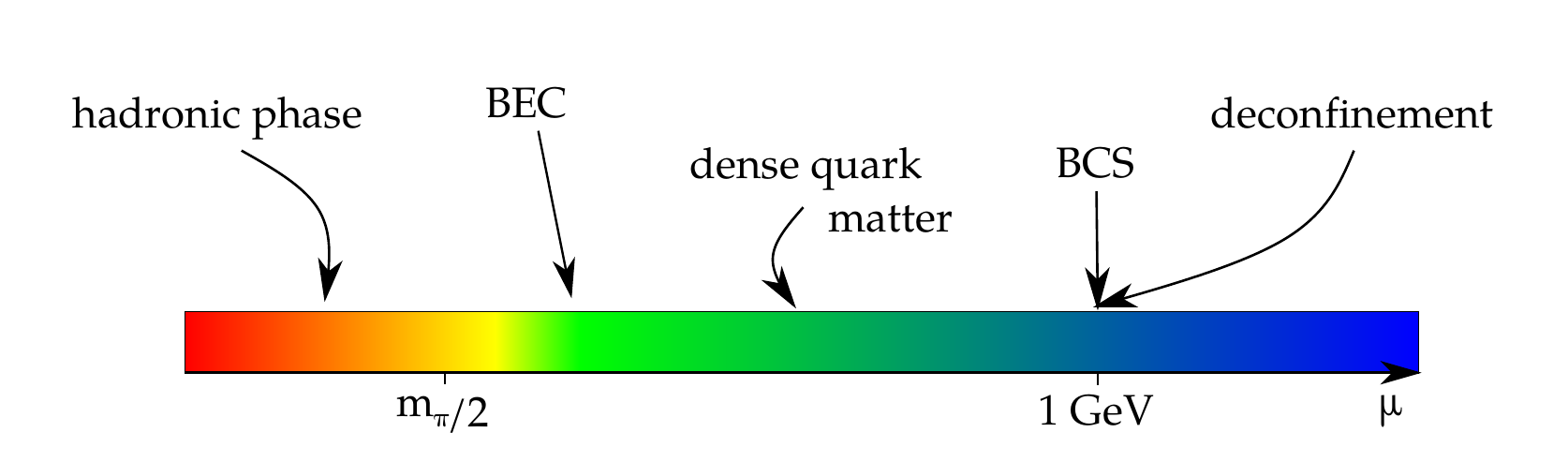}
    \caption{Schematic phase diagram of dense two-color QCD at low temperatures.}
    \label{fig:phase_diagram}
\end{figure}

At $\mu=m_{\pi}/2=370(20)$~MeV~($a\mu\simeq0.08$ in lattice units) there is a second order phase transition to a
phase where scalar diquarks form a Bose-Einstein condensate (BEC phase).
The order parameter of this transition is the diquark condensate $\langle q^T \bigl [ (C\gamma_5)\times \tau_2 \times \sigma_2 \bigr ] q \rangle$, 
where $C\gamma_5$ is the matrix which acts on Dirac indices and $\tau_2, \sigma_2$ are Pauli matrices which act on flavor and color indices of the quark field $q$. 
In the massless limit there is no chiral symmetry breaking, if the diquarks are condensed. 
However, for massive quarks  the chiral condensate is not zero. Instead it is proportional 
to the quark mass and 
decreases with increasing chemical potential. Let us note that dense QC$_2$D 
in the hadronic phase and 
the BEC phase was intensively studied within lattice simulations in a series of papers~\cite{Hands:1999md, Kogut:2001na, Kogut:2001if, Boz:2015ppa, Braguta:2016cpw, Holicki:2017psk}
where reasonable agreement with ChPT was observed.

In the ChPT the interactions between different degrees of freedom are accounted for by perturbation theory, so they are assumed to be weak and in addition the baryon density in the region of application is small. Together with the fact that in two-color QCD the diquarks are baryons this implies that the system is similar to a dilute baryon gas  at $\mu$  above $\mu\geq m_{\pi}/2$  but below the values corresponding to large density.

Enhancing the baryon density further, we proceed to dense matter, where the interactions between baryons cannot be accounted for within perturbation theory. This transition can 
be seen through the deviation of different observables from the predictions of ChPT. In our paper~\cite{Braguta:2016cpw}
this deviation was observed in the diquark condensate, the chiral condensate and the baryon density. 

At sufficiently large baryon density ($\mu \sim 1000~\mbox{MeV},~a\mu \sim 0.22$) some observables of the system under study can be described using Bardeen-Cooper-Schrieffer theory (BCS phase)\footnote{The properties of the BCS phase 
will be considered in a forthcoming study of ours.}. 
In particular, the baryon density is well described by the density of noninteracting 
fermions which occupy a Fermi sphere of radius $r_F=\mu$. In addition, the diquark 
condensate, which plays the role of a condensate of Cooper pairs, is proportional to 
the Fermi surface. In lattice simulation the BCS phase was observed in the following papers~\cite{Hands:2006ve, Hands:2010gd, Cotter:2012mb, Braguta:2016cpw} and we found that the transition from the BEC to the BCS phase is smooth~\cite{Braguta:2016cpw}. 

In addition to the transition to the BCS phase at $\mu \sim 1000$~MeV ($a\mu \sim 0.22$) there is the confinement/deconfinement transition in dense two-color QCD~\cite{Bornyakov:2017txe}. This transition manifests itself in a rise of the Polyakov loop and vanishing of the string tension. 
It was also found that after deconfinement is achieved, we observe a monotonous 
decrease of the spatial string tension $\sigma_s$ which ends up vanishing at $\mu_q \geq 2000$ MeV ($a\mu \geq 0.45$). It should be noted that the results of this study suggest that the confinement/deconfinement transition is rather smooth. 
The  Polyakov loop  results do not allow us to locate the transition region from the confinement to the deconfinement phase.
For this reason we consider here the transition region to be around $\mu = 1000$~MeV .  This value was found in our previous study \cite{Bornyakov:2017txe},
where it was determined by the condition  that the string tension, extracted from the Wilson loops, becomes zero within 
the uncertainty of the calculation. Thus throughout the paper we use the term "the confinement/deconfinement transition region" in the sense of vanishing of the string tension extracted  from the Wilson loops.

\section{The grand potential of a static quark-antiquark pair in dense quark matter}
\label{sec:grandpot}

In this section we are going to study the grand potential $\Omega_{\bar q q}(r, \mu)$ of a static quark-antiquark pair placed within a distance of $r$ into the dense medium. It can be represented in terms of the correlator of Polyakov loops 
\beq
\frac{\Omega_{\bar q q}(r, \mu)}{T} = - \log \langle \tilde \tr L_{\vec{x}} \tilde \tr L_{\vec{y}}^{\dagger}\rangle + c(\mu),\; r = |\vec{x} - \vec{y}|,
\label{omega_r_mu}
\eeq
where $\tilde \tr = \frac{1}{2} \tr$ and the Polyakov loop is given as the trace of a  product of gauge links in temporal direction $L_{\vec{x}} = \prod\limits_{\tau = 0}^{N_{\tau}-1} U_{\mu = 0}(\vec{x}, \tau).$  
The quantity $c(\mu)$ denotes a divergent renormalization constant, which is related to the self-energy of a quark or antiquark source. In the limit $r \to \infty$, the correlation between the Polyakov lines becomes negligible and the grand potential $\Omega_{\infty}(\mu)$ is given by the squared expectation value of the volume-averaged Polyakov loop, $\langle L \rangle = \langle N_s^{-3} \sum_{\vec{x}} \tilde \tr L_{\vec x}\rangle$:
\beq
\frac{\Omega_{\infty}(\mu)}{T} = \frac{1}{T}\lim\limits_{r \to \infty}\Omega_{\bar q q}(r, \mu) = -\log |\langle L \rangle|^2 + c(\mu).
\label{omega_r}
\eeq
To find the grand potentials from formulae (\ref{omega_r_mu}) and (\ref{omega_r}) one has to determine the renormalization constant $c(\mu)$.

In pure gauge theory the expectation value of the Polyakov line which is defined as
\beq
L^{ren}(\mu)=\exp {(-\Omega_{\infty}(\mu)/2T)},
\label{Polyakov_line_ren}
\eeq
is the order parameter of the confinement/deconfinement transition. 
In particular, $L^{ren}(\mu)$ vanishes in the confined phase, whereas it is non-zero in the deconfined phase.  
After inclusion of dynamical quarks in the simulations, the expectation value of the Polyakov line is no longer an order parameter.
However, one can interpret the $\Omega_{\infty}(\mu)/2$ as the grand potential of one quark or one antiquark in dense quark matter. 
Thus one may expect that in the confined phase $\Omega_{\infty}(\mu)$ is much larger than that in the deconfined phase.

Below we will also need the color-singlet grand potential $\Omega_1(r,\mu)$, which is defined as   
\beq
\frac{\Omega_1(r, \mu)}{T} = - \log\langle \tilde \tr (L_{\vec x} L_{\vec y}^{\dagger})\rangle + c^\prime(\mu).
\eeq
Notice that contrary to the color averaged grand potential $\Omega_{\bar q q}(r,\mu)$, 
the singlet one $\Omega_1(r,\mu)$ is not gauge invariant. So, in order
to calculate  $\Omega_1(r,\mu)$ we have to fix the gauge and we choose here conventionally the Coulomb gauge.

Both the color averaged and the color singlet grand potentials are calculated up to renormalization constants. Now let us define the relative normalization of these observables. It is clear that at sufficiently large spatial separation between quarks the relative orientation of charges in color space is not important due to screening. For this reason the authors of~\cite{Kaczmarek:2002mc, Kaczmarek:2004gv, Kaczmarek:2005ui} chose a relative normalization of the color averaged and color singlet free energies, such that they are identical at large distances. In our paper we are going to use the same relative normalization between the $\Omega_{\bar q q}(r, \mu)$ and $\Omega_1(r, \mu)$. 

For two-colors, the color averaged grand potential $\Omega_{\bar q q}(r, \mu)$ can be represented~\cite{Nadkarni:1986cz} through the 
color singlet $\Omega_1(r, \mu)$ and the color triplet grand potential $\Omega_3(r, \mu)$ as 
\beq
	\exp \left(-\frac{\Omega_{\bar q q}(r, \mu)}{T}\right) = \frac{1}{4} \exp \left(-\frac{\Omega_1(r, \mu)}{T}\right) + \frac{3}{4} \exp \left(-\frac{\Omega_3(r, \mu)}{T}\right).
\label{eq:exponents}
\eeq

Let us consider short distances ($r \mu \ll 1$), i.e. distances where the Debye screening can be neglected. 
In this limit the running of the coupling constant is determined by the scale $\sim 1/r$,  and the influence 
of the chemical potential on the running coupling can be neglected. The perturbative one-gluon exchange 
expression for the color singlet and the triplet grand potentials at short distances ($r \mu \ll 1$) can be written as
\beq
\Omega_1(r, \mu ) = -3 \Omega_3(r, \mu) + \mathcal{O}(g^4) = -\frac{g^2(r)}{8 \pi r} + \mathcal{O}(g^4).
\label{eq:relation}
\eeq

We have already discussed relative normalization between the grand potentials $\Omega_{\bar q q}(r, \mu)$ and $\Omega_1(r, \mu)$. Thus to renormalize the grand potentials it is sufficient to renormalize one of them. Let us consider $\Omega_1(r, \mu)$. 
To do this we use the procedure proposed in~\cite{Kaczmarek:2002mc, Kaczmarek:2004gv, Kaczmarek:2005ui},
adopted here for the calculation at finite density. The grand potential at finite temperature and chemical 
potential is defined as
\beq
\Omega_1(r, T, \mu) = U_1(r, T, \mu) - T S_1(r, T, \mu)-\mu N_1(r, T, \mu),
\label{grand_p}
\eeq
where function $U_1(r, T, \mu)$ is the internal energy, the function $S_1(r, T, \mu)$ is the entropy and the function $N_1(r, T, \mu)$ is the quark number density of a static color singlet quark-antiquark pair. From the above discussion it is clear that at short distances $r$ the grand potential 
does not depend neither on the chemical potential $\mu$ nor on the temperature $T$. This implies that at short distances 
the entropy $S_1 = - \partial \Omega_1 / \partial T$ and the quark number density $N_1 = - \partial \Omega_1 / \partial \mu$ are zero. 
It is also clear that at short distances the internal energy equals to the interaction potential in a quark-antiquark pair at zero temperature and density.
So, at short distances the grand potential $\Omega_1(r, T, \mu)$ coincides with the zero temperature and density 
potential $V(r)$ which is calculated in Appendix A. Similarly  to papers~\cite{Kaczmarek:2002mc, Kaczmarek:2004gv, Kaczmarek:2005ui}
we fix the renormalization constant $c'(\mu)$ through the matching condition for  $\Omega_1(r, \mu)$ at short distances to the 
short distance behavior of the interaction potential $V(r)$. 
The renormalization for the grand potential $\Omega_{\bar q q}(r, \mu)$ can be fixed using matching at large distances, $r$ where the color averaged and the color singlet grand potentials are expected to be identical.
Evidently, this procedure allows us to get rid of the divergent self-energy contributions and uniquely fixes the renormalization constants 
$c(\mu)$ and $c'(\mu)$.

\begin{figure}[t]
	\begin{minipage}[t]{0.48\textwidth}
    		\centering
		\includegraphics[width=1.00\textwidth]{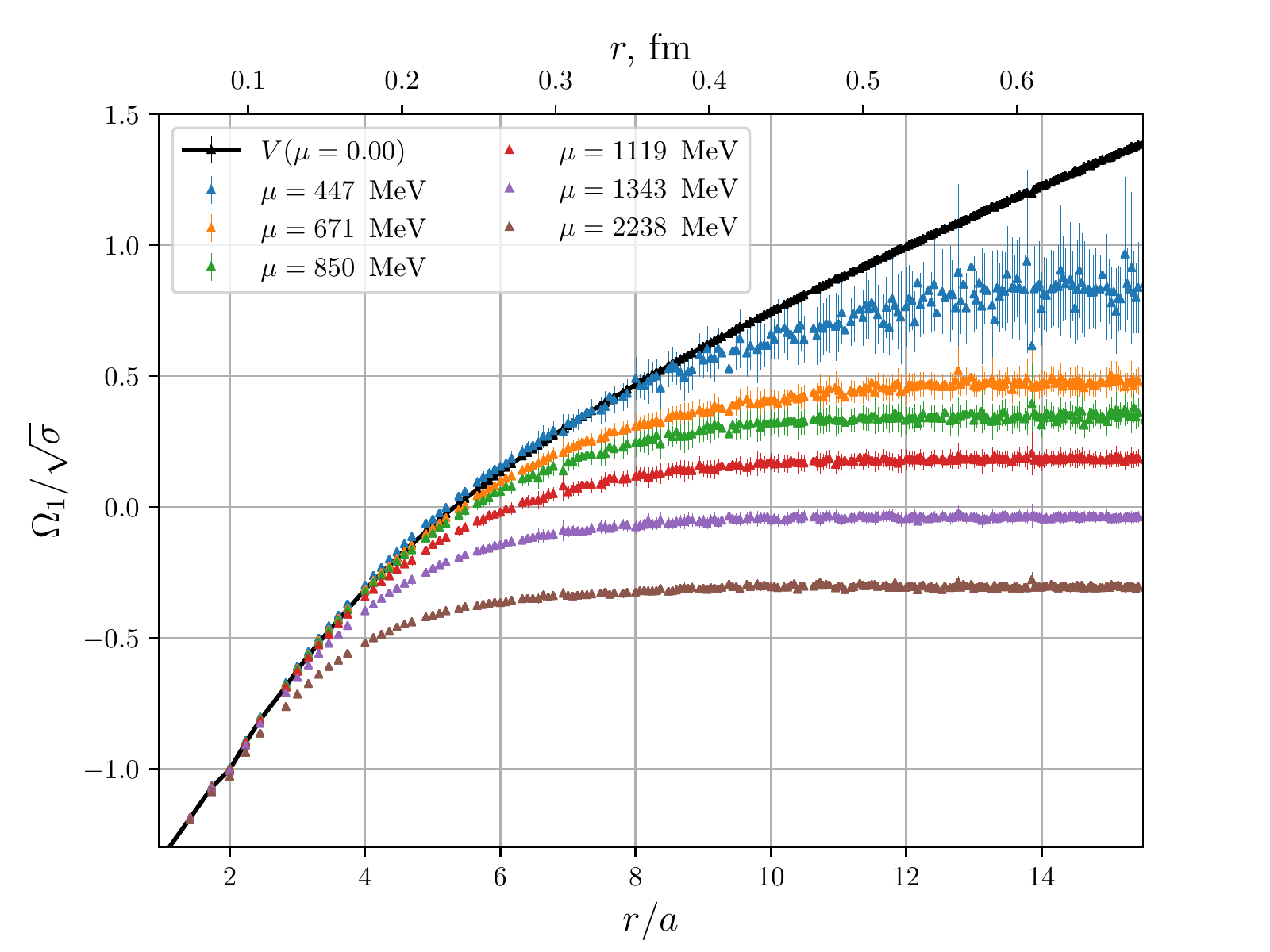}
		\caption{The color singlet grand potential as a function of distance for few values of the chemical potential under study. The black curve is the potential of a static quark-antiquark pair at zero density and temperature. Note the absence of a Coulombic small distance regime, due to smearing.}
		\label{fig:F1data}
	\end{minipage}
	\hfill
	\begin{minipage}[t]{0.48\textwidth}
		\centering
		\includegraphics[width=1.00\textwidth]{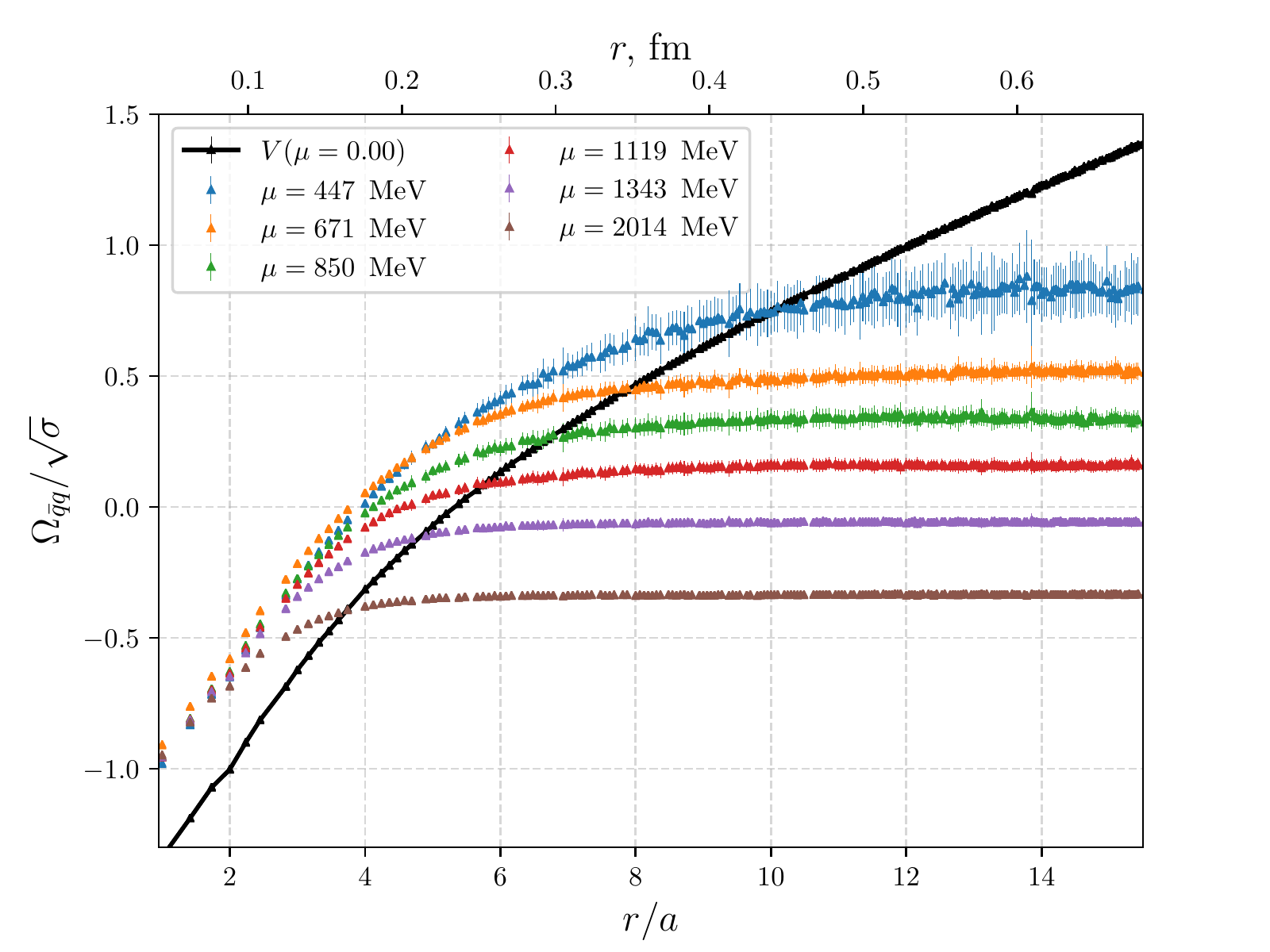}
		\caption{The $\Omega_{\bar q q}$ as a function of distance for few values of the chemical potential under study. The black curve is the potential of the static quark-antiquark pair at zero density and temperature.Note the absence of a Coulombic small distance regime, due to smearing. }
		\label{fig:Fdata}
	\end{minipage}
\end{figure}

\begin{figure}[t]
	\centering
	\includegraphics[width=0.75\textwidth]{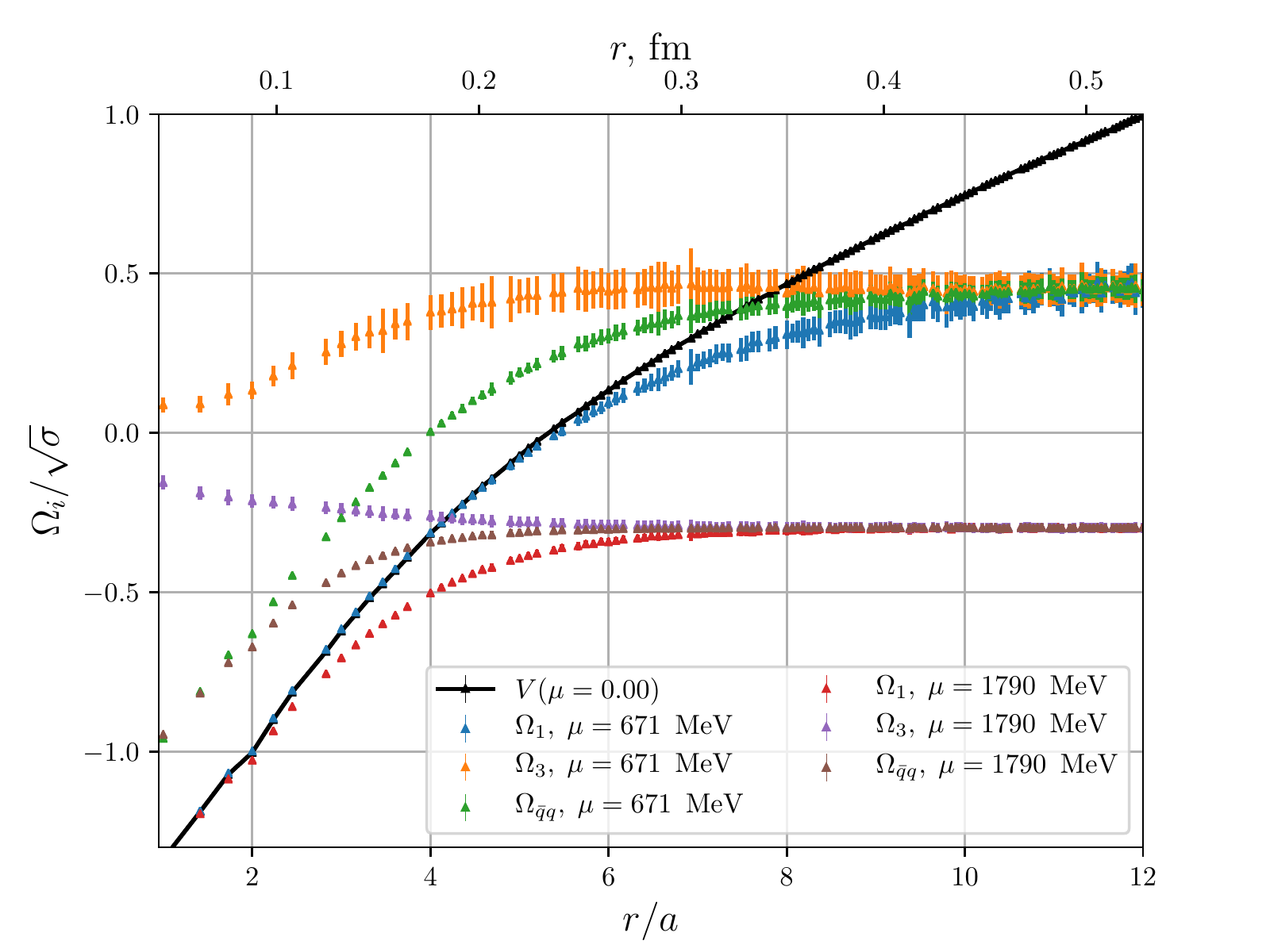}
	\caption{The singlet, triplet and the color averaged grand potentials for two values of the chemical potential: $\mu=671$~MeV and $\mu=1790$~MeV.}
	\label{fig:F1F3F}
\end{figure}

Having gone through the renormalization procedure we are ready to present the results of the calculation of the renormalized 
grand potentials $\Omega_{\bar q q}(r, \mu)$, $\Omega_1(r, \mu)$, $\Omega_3(r, \mu)$. 
In figure~\ref{fig:F1data} we plot the renormalized $\Omega_1(r,\mu)$ as a function of distance for different values of the chemical potential. 
In figure~\ref{fig:Fdata} we plot the  grand potential 
$\Omega_{\bar q q}(r, \mu)$. 
To get an idea how the $\Omega_{\bar q q}(r, \mu)$, $\Omega_1(r, \mu)$, $\Omega_3(r, \mu)$ look in one figure 
we plot figure~\ref{fig:F1F3F} where these potentials are shown for the values $\mu=671$~MeV 
and $\mu=1790$~MeV. 

\begin{figure}[t]
	\begin{minipage}[t]{0.48\textwidth}
     \centering
     \includegraphics[scale=0.49]{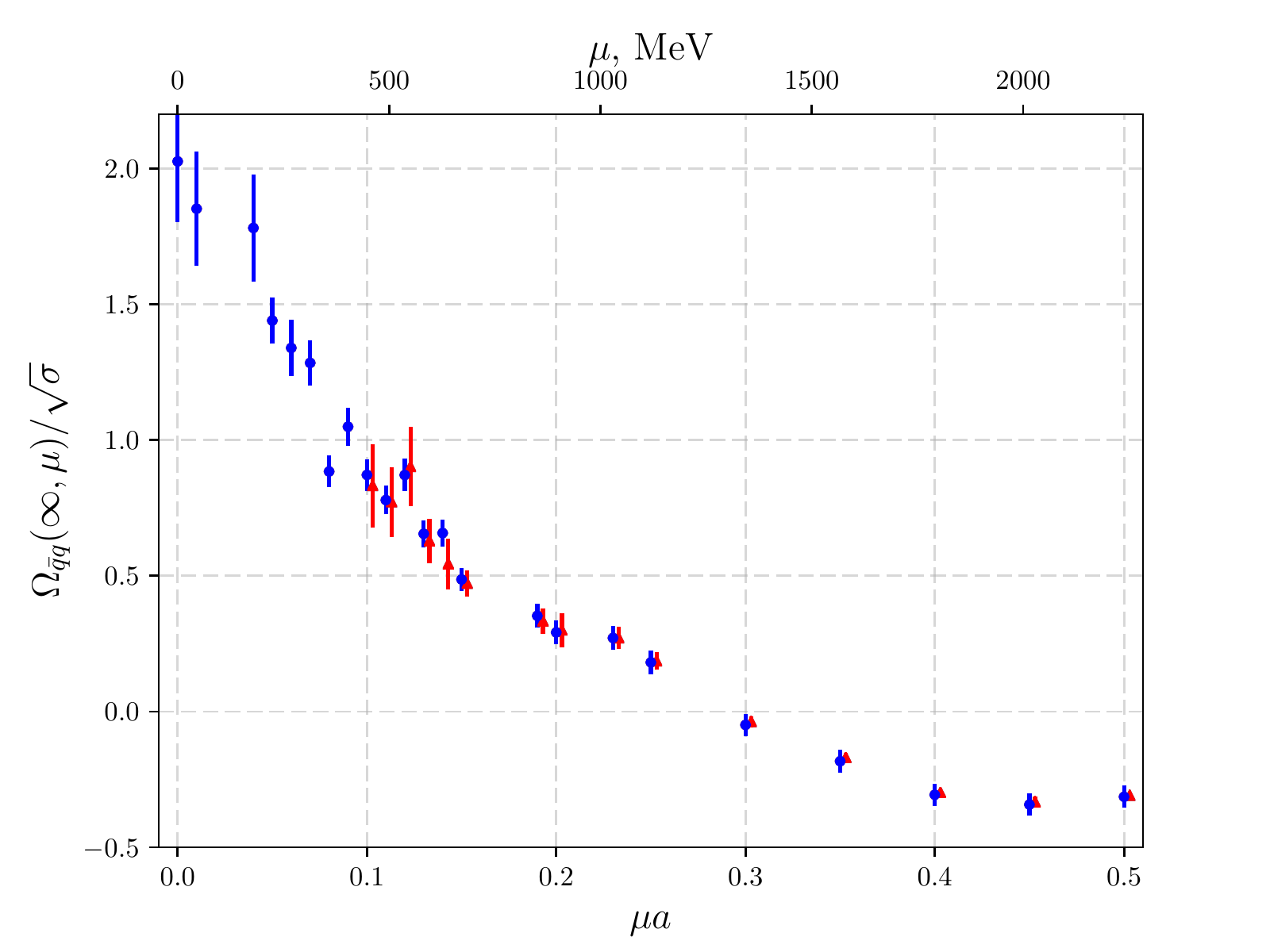}
     \caption{The renormalized color singlet grand potential $\Omega_1(\infty, \mu)$ as a function of $\mu$. The red tringles correspond to the $\Omega_1(\infty, \mu)$ extracted from the renormalized correlators of Polyakov loops. The blue circles correspond to the $\Omega_1(\infty, \mu)$ extracted from the average Polyakov loops measured on the lattice and renormalized according to formula (\ref{Lren}).}
     \label{fig:F1_inf}
   \end{minipage}
   \hfill
   \begin{minipage}[t]{0.48\textwidth}
     \centering
     \includegraphics[scale=0.49]{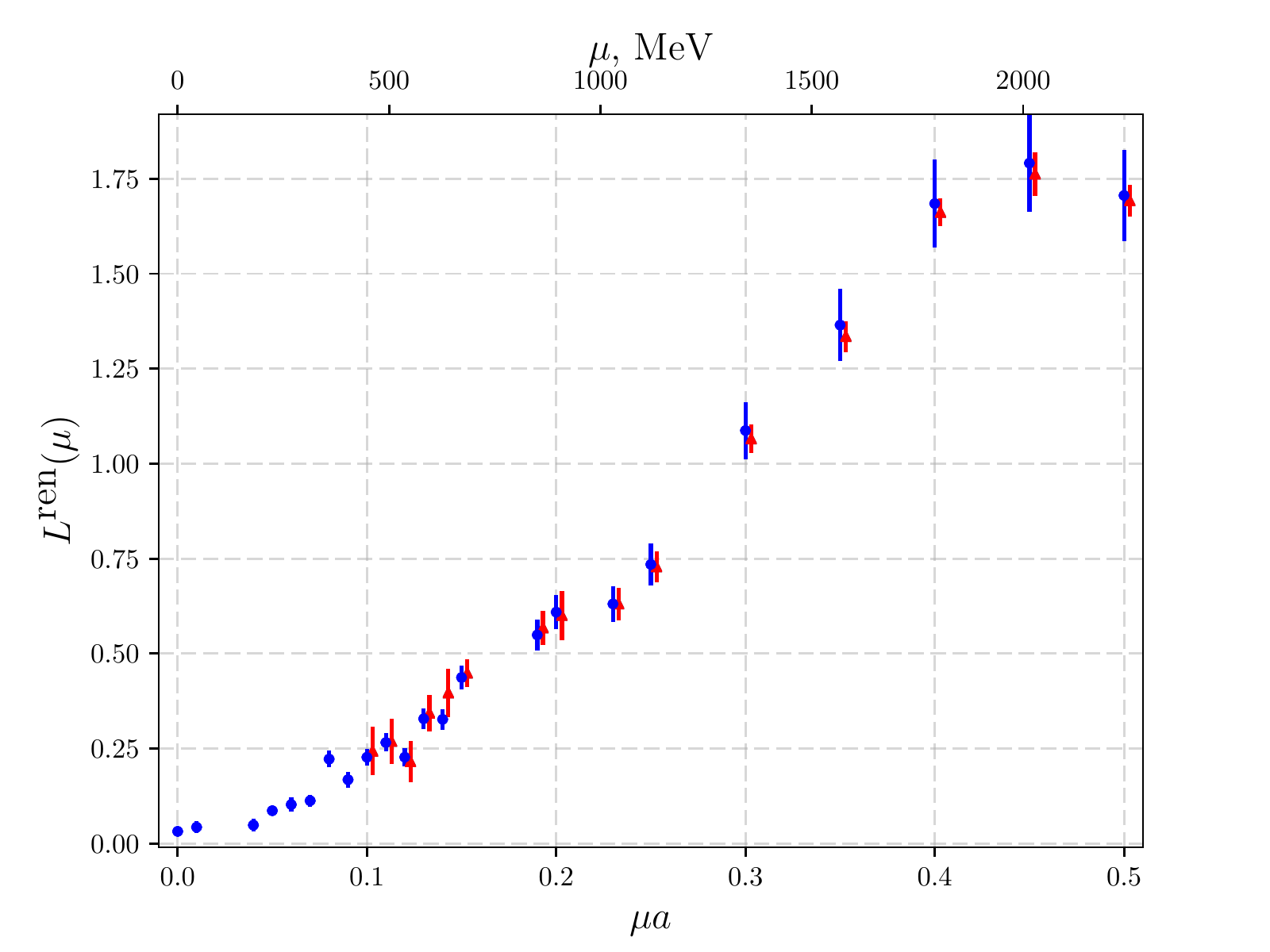}
     \caption{The renormalized Polyakov loops $L^{ren}(\mu)$ as a function of $\mu$. The red tringles correspond to the $L^{ren}(\mu)$ extracted from the renormalized correlators of Polyakov loops. The blue circles correspond to the average Polyakov loops measured on the lattice and renormalized according to formula (\ref{Lren}).}
     \label{fig:F1_inf_exp}
   \end{minipage}
\end{figure}

The grand potential of a single quark/antiquark in quark matter and the Polyakov loop are important quantities in two-color QCD. 
After the renormalization these observables can be extracted from the Polyakov loop correlator at large distances. 
In the calculation we take $\Omega_1(\infty, \mu)=\Omega_1(L_s/2, \mu)$ and calculate the renormalized Polyakov loop 
applying formula (\ref{Polyakov_line_ren}). Notice that for the calculation it is important that 
the grand potential extracted from the Polyakov loop correlator goes to a plateau value. 
In the confined, phase the plateau in the grand potential is due to string breaking, which takes place 
at large distance. Due to the relatively small spatial lattice size we can observe the string breaking
only for sufficiently large chemical potential ($\mu > 440$~MeV). For this reason 
the calculation of the $\Omega_1(\infty, \mu)$ and $L^{ren}(\mu)$ based on the renormalized 
correlator of the Polyakov loops can be carried out for $\mu > 440$~MeV. The results for $\Omega_1(\infty, \mu)$ and $L^{ren}(\mu)$
are shown in figure~\ref{fig:F1_inf} and in figure~\ref{fig:F1_inf_exp} by red triangles.

The renormalization of the $\Omega_1(\infty, \mu)$ and the Polyakov line can be carried out through 
the measurement of the latter on the lattice. In this case it is possible to find both observables 
for all values of the chemical potential under study. To calculate the Polyakov loop 
we conduct one step of the HYP smearing. The Polyakov loop is renormalized according to
\beq
L^{ren}(\mu)=L^{bare}(\mu) \frac {L^{ren}(\mu=1030~\mbox{MeV})} {L^{bare}(\mu=1030~\mbox{MeV})},
\label{Lren}
\eeq
where $L^{ren}(\mu=1030~\mbox{MeV})$
is the Polyakov loop measured in the previuos approach\footnote{The point $\mu=1030~$MeV was chosen since here we have large statistics and, therefore, rather good accuracy in the calculation of the $L^{ren}(\mu=1030~\mbox{MeV})$.}, 
and $L^{bare}(\mu)$ is the bare Polyakov loop measured on the lattice. Similarly to the renormalization of the 
correlators of the Polyakov loops, the approach based on (\ref{Lren}) gets rid of infinite ultraviolet divergence and 
uniquely fixes the renormalization. 

Having calculated the renormalized Polyakov line, we can find the $\Omega_1(\infty, \mu)$ using formula (\ref{Polyakov_line_ren}).
The results for $\Omega_1(\infty, \mu)$ and $L^{ren}(\mu)$
are shown in figure~\ref{fig:F1_inf} and in figure~\ref{fig:F1_inf_exp} by blue circles. 
From these figures one sees that both approaches to the calculation of the $\Omega_1(\infty, \mu)$ and $L^{ren}(\mu)$
are in agreement with each other. 

Here a few comments are in order:  the measurement of the Polyakov loop correlation functions 
in this section was carried out at $\lambda=0.00075$.
As was discussed above, in order to check the $\lambda$-dependence of our results we carried out a similar study 
at $a\mu=0.0, 0.1, 0.2, 0.3, 0.4$ and $\lambda=0.0005, 0.001$. We found that with the 
exception of the chemical potential $a\mu=0.4$
the results obtained with different values of the $\lambda$ parameter are in agreement with each other within the uncertainty of the calculation. 
For the chemical potential $a\mu=0.4$ the results obtained at different $\lambda$ deviate from each other by around $2\times \sigma$.
From this fact we conclude that the $\lambda$-dependence of our results is weak.

The confinement/deconfinement transition at finite temperature manifests itself in an increasing value of the Polyakov loop and its rise may become quite rapid in the transition region~\cite{Kaczmarek:2005ui}. A similar 
behaviour can be seen from figure~\ref{fig:F1_inf_exp}, where the confinement/deconfinement transition is observed through the rise of the Polyakov line. 
At the same time from this figure we don't see any specific region in the chemical potential where the rise of the Polyakov line 
is dramatically different as compared to other regions. 
This observation corroborates our previous finding that the finite density confinement/deconfinement transition 
in two-color QCD is rather smooth. 
\footnote {Our data do not allow us to confirm or exclude existence of an inflection point at some value of the chemical
potential similar to the inflection point in the temperature dependence of the Polyakov
loop. The search of possible inflection point in dense matter requires better accuracy and
more data points as compared to the data available in this study.}

Let us now pay attention to the region $\mu>2000~$MeV. In this region the Polyakov loop/grand potential reaches a maximum/minimum and then drops/rises. Below it will be shown that the region $\mu>2000~$MeV differs from the region $\mu < 2000~$MeV not only for the Polyakov loop 
but also the grand potential. In turn also derived  observables, such as the screening length $R_{sc}$, the Debye mass and effective coupling constant show a distinctive behavior.

At the moment we do not fully understand the physics, which is responsible for this behavior. One possibility is that the value of the chemical potential 
$\mu \sim 2000~$MeV is exceptional since there is nonzero spatial string tension in the region $\mu<2000~$MeV whereas 
the spatial string tension is zero for $\mu > 2000~$MeV. This might imply that the point $\mu \sim 2000~$MeV separates 
systems with and without magnetic screening. However a definite answer to this hypothesis requires further study. 

From figure~\ref{fig:F1_inf} one sees that the potential $\Omega_{\bar q q}(r,\mu)$ changes its sign at $\mu \sim 1300~$MeV, which at first sight may be unexpected. However, let us recall that $\Omega_{\bar q q}(r,\mu)$ is not a grand potential of the whole system.
On the contrary, it is the difference of the grand potential of dense quark matter with a static quark-antiquark pair 
and dense quark matter without a static quark-antiquark pair. So, $\Omega_{\bar q q}(r,\mu)$ in our context is the 
additional grand potential due to the introduction of the quark-antiquark pair to quark matter. Now figure~\ref{fig:F1_inf}
can be interpreted as follows: introducing a static quark-antiquark pair increases the grand potential of the system
for $\mu<1300~$MeV and decreases the grand potential of the system for $\mu>1300~$MeV. 
An explanation of this fact will be presented in the next section. 

The authors of~\cite{Cotter:2012mb} studied QC$_2$D with $N_f=2$ quarks within lattice simulation with dynamical Wilson fermions. 
In particular, they measured the Polyakov loop as a function of the chemical potential and observed the following behavior:
It remains zero up to $a\mu \sim 0.75$ and then quickly rises. In the region $a \mu>1$, due to the saturation the quark degrees of 
freedom, quarks are no longer dynamical and the theory becomes quenched QCD and exhibits confinement, i.e. the Polyakov loop goes to zero for $a \mu>1$. 

Further measurement of the string tension carried out in~\cite{Boz:2013rca} did not confirm the presence of a confinement/deconfinement transition
and the decrease of the string tension with chemical potential. Although the behavior of the Polyakov line in figure~\ref{fig:F1_inf_exp} 
seems similar to that obtained in~\cite{Cotter:2012mb}, we believe that this behavior can be explained as a lattice artifact of Wilson fermions. 
In order to explain the results of~\cite{Cotter:2012mb}, let us recall that in Wilson fermions one has one light quark and 
15 heavy quark species with masses $\sim 1/a$. If the chemical potential is $a\mu \sim 1$ the heavy quarks are not suppressed any longer and
additional color degrees of freedom are released to the system under study. We believe that this mechanism is responsible for the 
rise of the Polyakov line observed in~\cite{Cotter:2012mb}. Notice that this mechanism does not work for staggered quarks as used in 
our paper, since there are no heavy species. In addition we observe a considerable rise of the Polyakov loop already at $a\mu \sim 0.2$. 
The decrease of the Polyakov line in figure~\ref{fig:F1_inf_exp} hence cannot be attributed to the saturation since 
it starts at $a\mu \sim 0.4$, which is rather far away from the saturation in staggered fermions~\cite{Braguta:2016cpw}.  

\section{The quark number and internal energy induced by a static quark-antiquark pair}
\label{sec:quarknum}

The authors of~\cite{Kaczmarek:2005gi} calculated the free energy of a
static quark-antiquark pair in QCD at finite temperature. 
In addition they calculated the entropy of the QCD medium in the presence of a static-quark antiquark pair. In dense quark matter 
there is in addition a contribution of the entropy to the grand potential (\ref{grand_p}). However, 
this contribution is not important at low temperature. What becomes important in dense quark matter is
the quark number induced by the static quark antiquark pair $N(r)$. This quantity can be 
calculated as follows
\beq
N(r, \mu)=- \frac {\partial \Omega(r,\mu)} {\partial \mu}.
\eeq
Notice that the $N(r, \mu)$ is the quark number which arise in the system due to the introduction of the quark-antiquark pair to the dense matter. So, it is the difference of the quark number with and without static quark-antiquark pair. 

\begin{figure}[t]
   \begin{minipage}[t]{0.48\textwidth}
     \centering
     \includegraphics[scale=0.5]{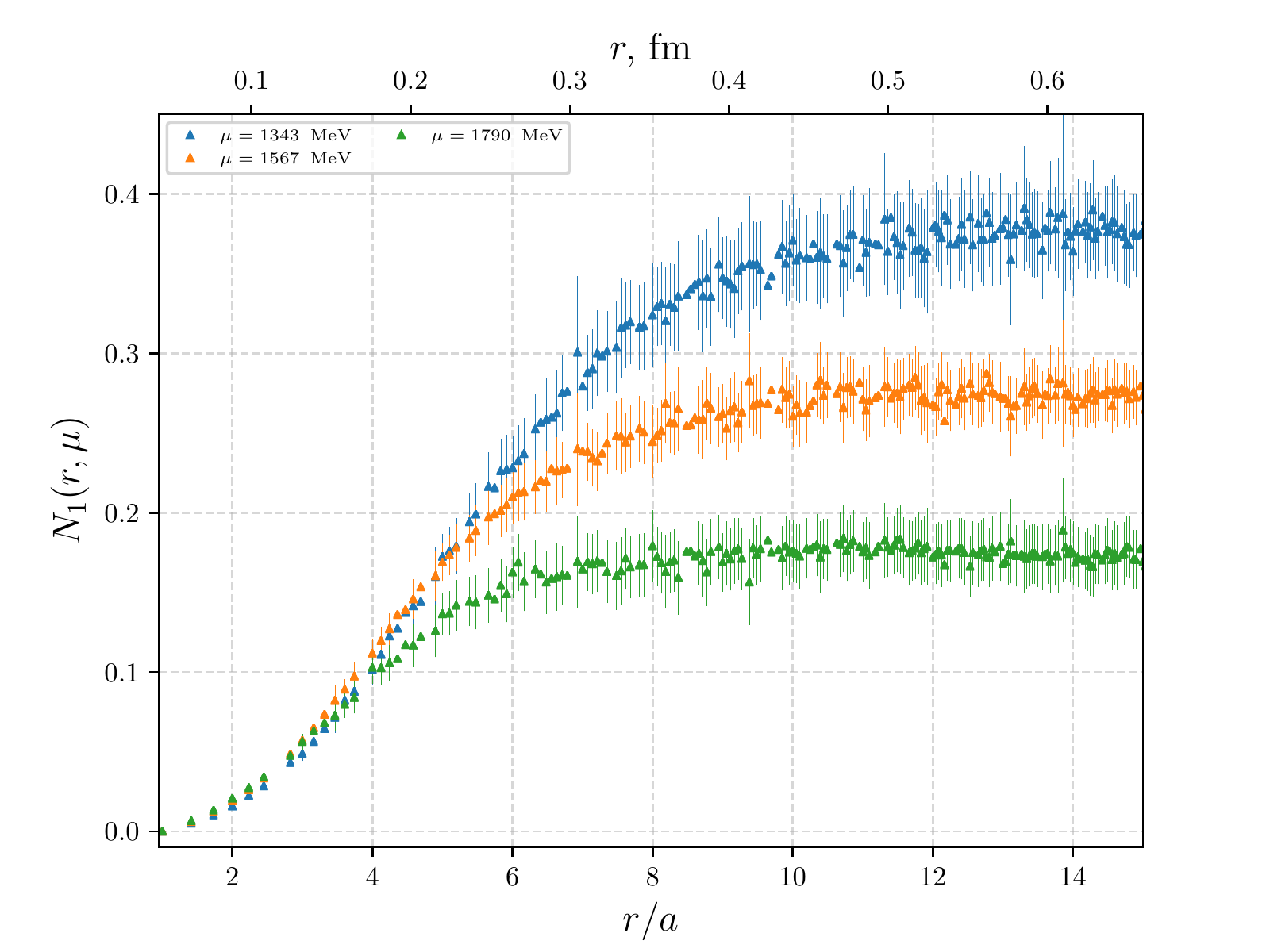}
     \caption{The quark number $N_1(r,\mu)$ induced by a static quark-antiquark pair as a function of distance for few values of chemical potential.}
     \label{fig:N_1}
   \end{minipage}
   \hfill
	\begin{minipage}[t]{0.48\textwidth}
     \centering
\includegraphics[scale=0.5]{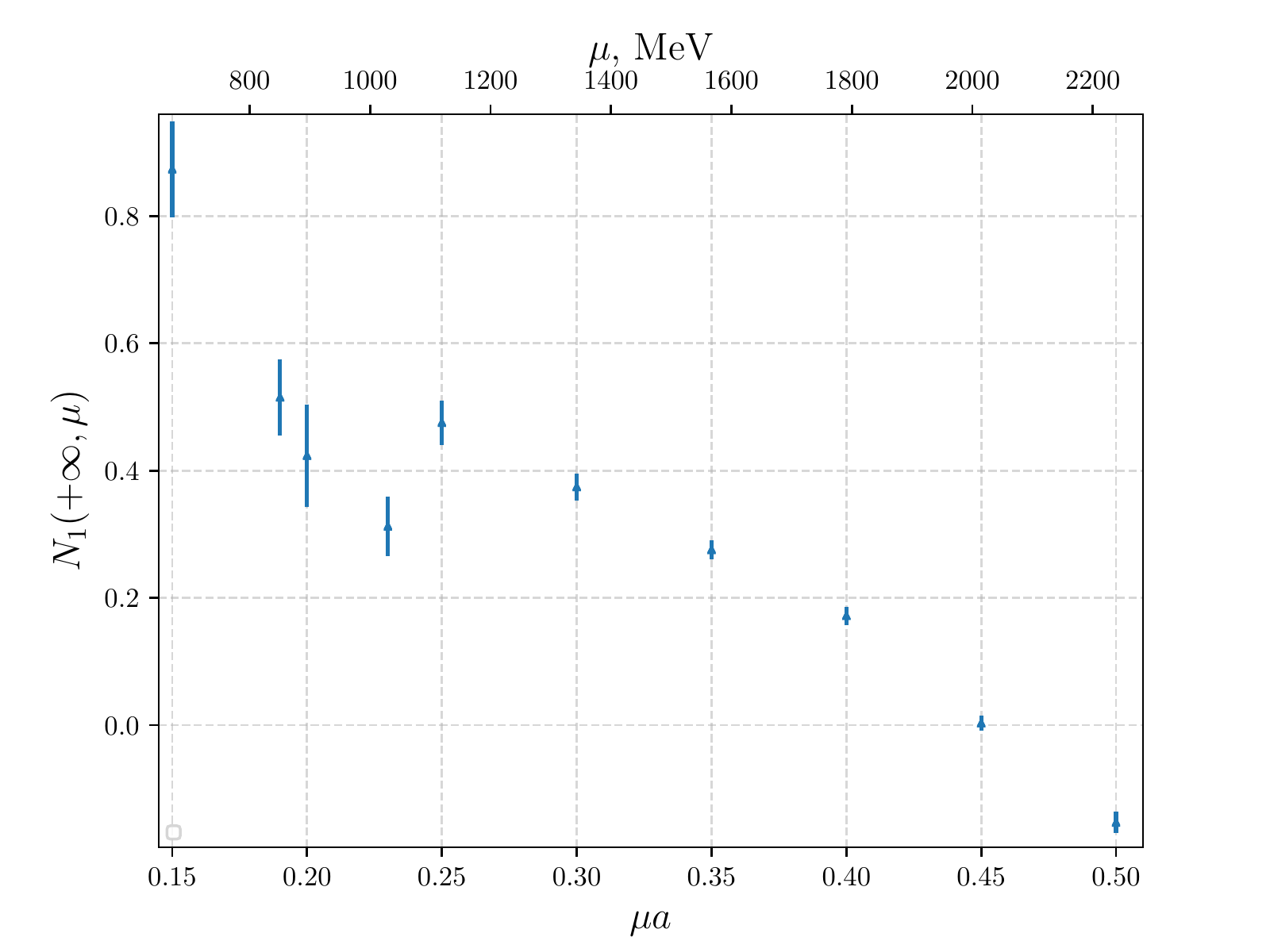}
\caption{The quark number $N_1(r,\mu)$ at large distance as a function of chemical potential.}
     \label{fig:N1_inf}
   \end{minipage}
\end{figure}

In this paper we calculate the $N_1(r, \mu)$ for the color singlet grand potential. Similar 
observables can be calculated for the color averaged and color triplet grand potentials. 
To find the $N_1(r, \mu)$ we determine the derivative of the grand potential over chemical potential through the finite difference approximation. 
The results of the calculation of the $N_1(r, \mu)$ for a few values of the chemical potential are shown in figure~\ref{fig:N_1}.
We found that the smaller the chemical potential, the larger uncertainty of the calculation in the $N_1(r, \mu)$.
For this reason we show only for $N_1(r, \mu)$ for sufficiently large chemical potentials. 

From figure~\ref{fig:N_1} one sees that $N_1(r, \mu)$ is rising from zero at short distances to some plateau value $N_1(\infty, \mu)$, which is an important observable, since it is proportional to the derivative of the Polyakov loop of a single quark/antiquark
in a dense medium over the chemical potential. One might expect that at the critical chemical potential where the confinement/deconfinement 
phase transition takes place there is an inflection point of the Polyakov loop. For this reason the confinement/deconfinement 
phase transition might manifest itself in the maximum of $N_1(\infty, \mu)$. For the same reason the authors of~\cite{Kaczmarek:2005gi} observed a maximum of the entropy at the critical temperature. Our results for 
$N_1(\infty, \mu)$ are shown in figure~\ref{fig:N1_inf}. Unfortunately due to large uncertainties of the calculation we are unable to locate this maximum.

\begin{figure}[t]
	\centering	
	\includegraphics[scale=0.7]{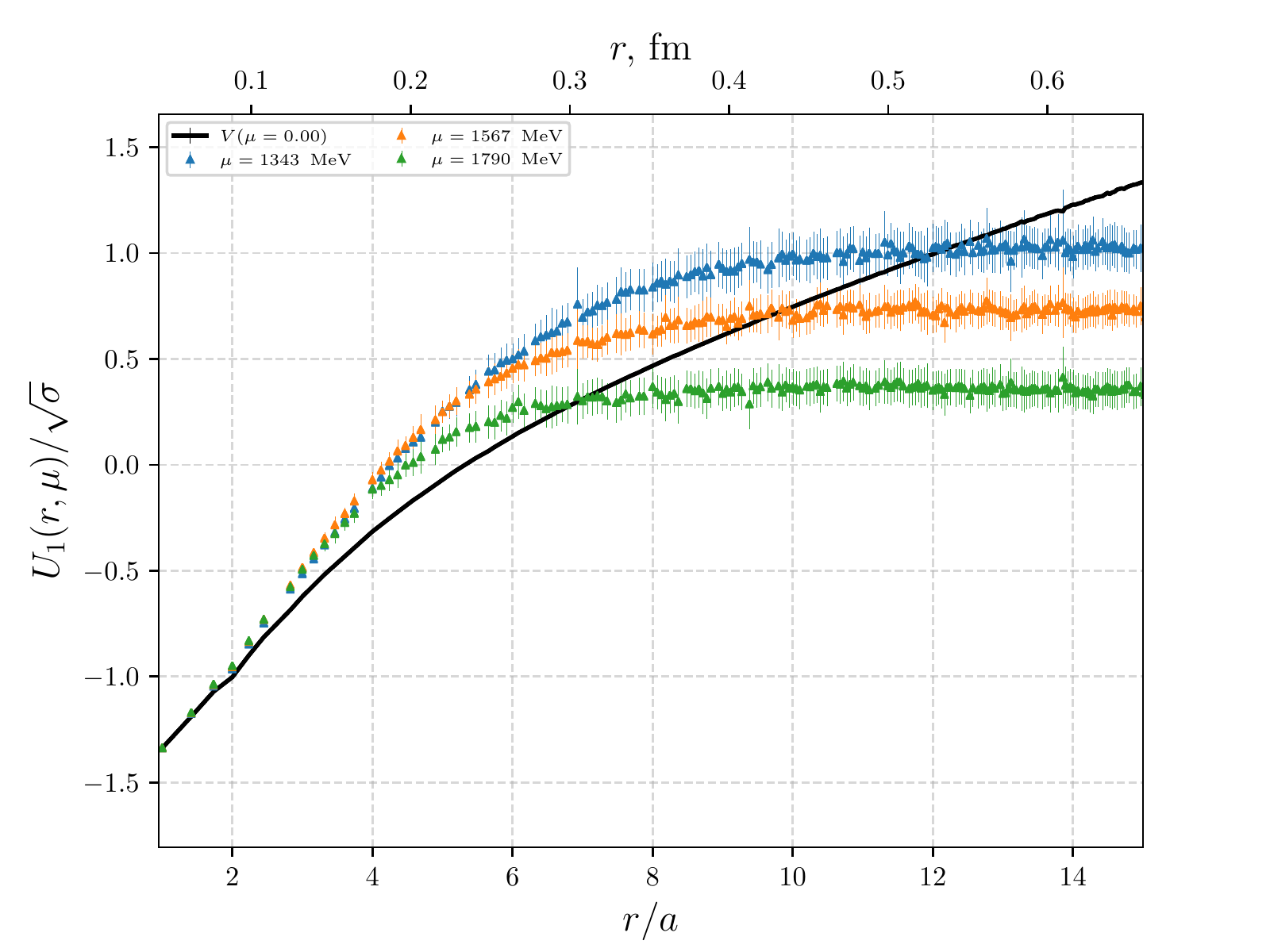}
	\caption{The internal energy of a static quark-antiquark pair as a function of the distance for few values of the chemical potential under study. The black curve is the potential of static quark-antiquark pair at zero density and temperature.}
	\label{fig:U_1}
\end{figure}

If we ignore the entropy contribution to the grand potential, which is small in cold dense matter, one can calculate also the internal $U(r,\mu)$ 
energy using the formula
\beq
U(r,\mu)=\Omega(r,\mu) + \mu N(r,\mu). 
\eeq
In this paper we calculate the $U_1(r, \mu)$ for the color singlet grand potential, the result of which is shown in figure~\ref{fig:U_1}. 

\section{String breaking in dense quark matter}
\label{sec:stringbreak}

Let us consider again figure~\ref{fig:F1data} and figure~\ref{fig:Fdata}. We have already mentioned 
that in dense QC$_2$D the confinement/deconfinement transition takes place at $\mu\sim 1000~$MeV. Despite 
this fact from figure~\ref{fig:Fdata} we see that $\Omega_{\bar q q}(r,\mu)$ reaches the plateau already at $\mu=447~$MeV. 
This happens because of the string breaking phenomenon, which for $\mu=447~$MeV takes place at $r\sim 0.5~$fm. 
Of course string breaking occurs also for smaller chemical potentials, but we do not observe it, as it takes place beyond our spatial lattice size. From figure~\ref{fig:F1data} one also sees that the larger 
the chemical potential, the smaller the distance at which the string breaking takes place. Let us consider this phenomenon more quantitatively
\footnote{At sufficiently large values of the chemical potential the screening properties of
the dense medium are described by the Debye screening phenomenon rather than by the string breaking.
In this section we consider the chemical potential values corresponding to nonzero string tension extracted from the Wilson loops \cite{Bornyakov:2017txe}.}. 

\begin{figure}[t]
	\centering
	\includegraphics[scale=0.725]{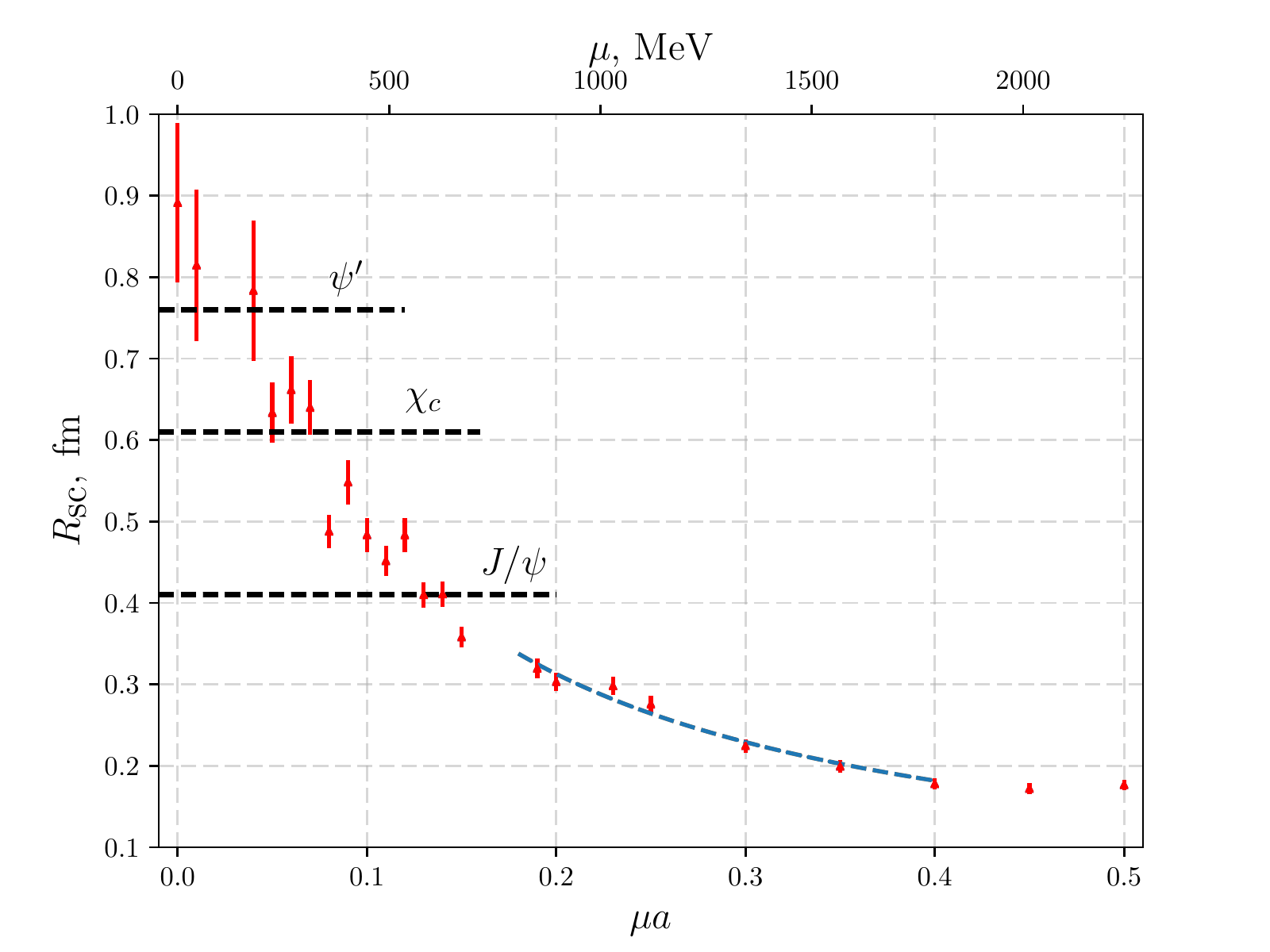}
	\caption{The screening length calculated from equation (\ref{r_screening}) as a function of chemical potential. Black dashed lines represent mean squared radii $\sqrt{\langle r^2\rangle}$ of charmonia calculated in Appendix B. The blue dashed line is the description of the screening length $R_{sc}$ by the Debye screening formula (\ref{rsc_debye}).}
	\label{fig:r_screening}
\end{figure}

To study the string breaking phenomenon we introduce the screening length $R_{sc}$ which can be calculated 
from the solution of the equation~\cite{Kaczmarek:2002mc}
\beq
V_{\mu=0}(R_{sc})=\Omega_{\bar q q} (\infty,\mu),
\label{r_screening}
\eeq
where $V_{\mu=0}(r)$ is the static potential at zero density. For $\Omega_{\bar q q} (\infty,\mu)$
we take the grand potential calculated from the renormalized Polyakov loop measured on the lattice(see figure~\ref{fig:F1_inf}).
The results of this calculation are shown in figure~\ref{fig:r_screening}. This plot tells us that the larger the chemical 
potential the smaller the string breaking distance. 

In order to understand this behaviour, let us recall that in three-color QCD the string breaking phenomenon 
can be explained by the possibility to break the string between static quarks by a quark-antiquark pair created 
from vacuum. If the length of the string is larger than the critical one it becomes energetically favorable 
to break the string and form two heavy-light meson instead of increasing the length of the string. 

In dense two-color QCD in addition to the possibility to break the string by quark-antiquark pairs it becomes 
possible to break the string by two quarks. As the result of this phenomenon, after the string has been broken, one ends up with a heavy-light meson
and one heavy-light diquark. Due to confinement, the two quarks have to be extracted from some hadron.
The two-color baryon -- the scalar diquark is a good candidate for such a hadron. Indeed at nonzero $\mu$ the scalar 
diquark is a lightest state in the system. At $\mu>m_{\pi}/2$ there is condesation of the scalar diquarks, 
so the two quarks can be extracted from the diquark condensate, which does not require any energy. This picture is supported at large $\mu$
in the BCS phase, where one has a Fermi sphere with radius $\mu$. Evidently one cannot break the string 
by taking two quarks deep inside the Fermi sphere, since in that case, the quarks which break the string due to the interactions have to move from one point of the Fermi sphere to some other point inside the Fermi sphere. However, all points inside the Fermi sphere are occupied.
So, the only possibility to break the string is to take two quarks close to the Fermi surface. 
In the confined phase, quarks on the Fermi surface are condensed as diquarks. Thus we again 
confirm the picture that two quarks, which break the string between a quark-antiquark pair, 
can be taken from the available diquarks. 

Further, let us consider the following model: if one diquark penetrates inside the string,
it breaks the string with some probability $\omega$. For the density of diquarks $n$, the string 
length $R$ and the transverse area $S$ the number of diquarks inside the string is $n \times R \times S$. 
If the string breaking events are independent, the total probability to break the string $P \simeq \omega \times n \times R \times S$. 
The condition for the string breaking is $P \sim 1$. From last statement we conclude that $R_{sc} \sim 1/n$. 
Finally the larger the chemical potential, the larger the condensate and the density of diquarks, which explains 
why $R_{sc}$ is decreasing with the $\mu$. 

If one increases the chemical potential then at some density $R_{sc}$ becomes so small 
that the string cannot be created. 
I.e. at the instant of creation it will be immediately broken by the 
two-color baryons -- diquarks. This is our hypothesis of the deconfinement mechanism in two-color 
dense quark matter. It is not clear how to find unambiguously the distance at which the 
string ceases to be stable. From the interaction potential at zero density (see figure~\ref{fig:F1data}) it is 
found that this happens in the region $r\in (0.2,0.3)~$fm. Using figure~\ref{fig:r_screening} one can 
infer that the interactions in this interval are screened, as the chemical potential lie within $\mu \in (900, 1300)$ and which agrees with the position of the confinement/deconfinement transition. We believe that 
this fact confirms our hypothesis. 

For a chemical potential larger than the $\mu\sim 1300~$MeV($a\mu\sim 0.3$) the $R_{sc}$ is smaller 
than $0.2~$fm. At such small distances the entropy $S = - \partial \Omega / \partial T$, the quark number density $N = - \partial \Omega_1 / \partial \mu$
are small and the grand potential is mainly determined by the interaction potential at zero temperature and density. 
At the same time the renormalized interaction potential (see figure~\ref{fig:F1data}) is negative at distances $r<0.2~$fm. 
For this reason the grand potential of one quark in dense quark matter becomes negative, which was observed in the previous section. 

In addition to the $R_{sc}$ in figure~\ref{fig:r_screening} we plot the average heavy quarkonia $J/\Psi, \chi_c, \psi'$ 
radii which where estimated in Appendix B within a simple potential model. It is clear that if the screening length is close to the heavy quarkonium 
radius this state is considerably modified by dense quark matter. From figure~\ref{fig:r_screening} one sees that
the heaviest state the $\psi'$ due to its rather large radius should be considerably modified at nonzero density before 
the transition to BEC phase. The $\chi_c$ meson will instead be modified in the BEC phase. Finally 
we predict that the $J/\Psi$ meson will be modified in dense quark matter but below the deconfiment transition. Notice, however,
that if the radius of a charmonium equals to the $R_{sc}$ at some density $n_0$, the dissociation of this charmonium 
takes place at densities larger than $n_0$.

A more detailed study of quarkonium dissociation in two-color dense quark matter will be presented in a future study. In particular the question of the presence of an imaginary part in the interquark potential at finite density, which may further destabilize the bound states will be carefully investigated.

\section{Debye screening in dense quark matter}
\label{sec:DebyeScreen}

In the region $\mu>900~$MeV the system under study transitions from the confined to the deconfined phase. In the deconfined phase the contribution of the string is markedly reduced and one may attempt to describe $R_{sc}$ in a dense quark-gluon plasma via an analogy with the Abelian theory, i.e. purely Coulombic Debye screening. 

\begin{figure}[b]
	\centering	
	\includegraphics[scale=0.65]{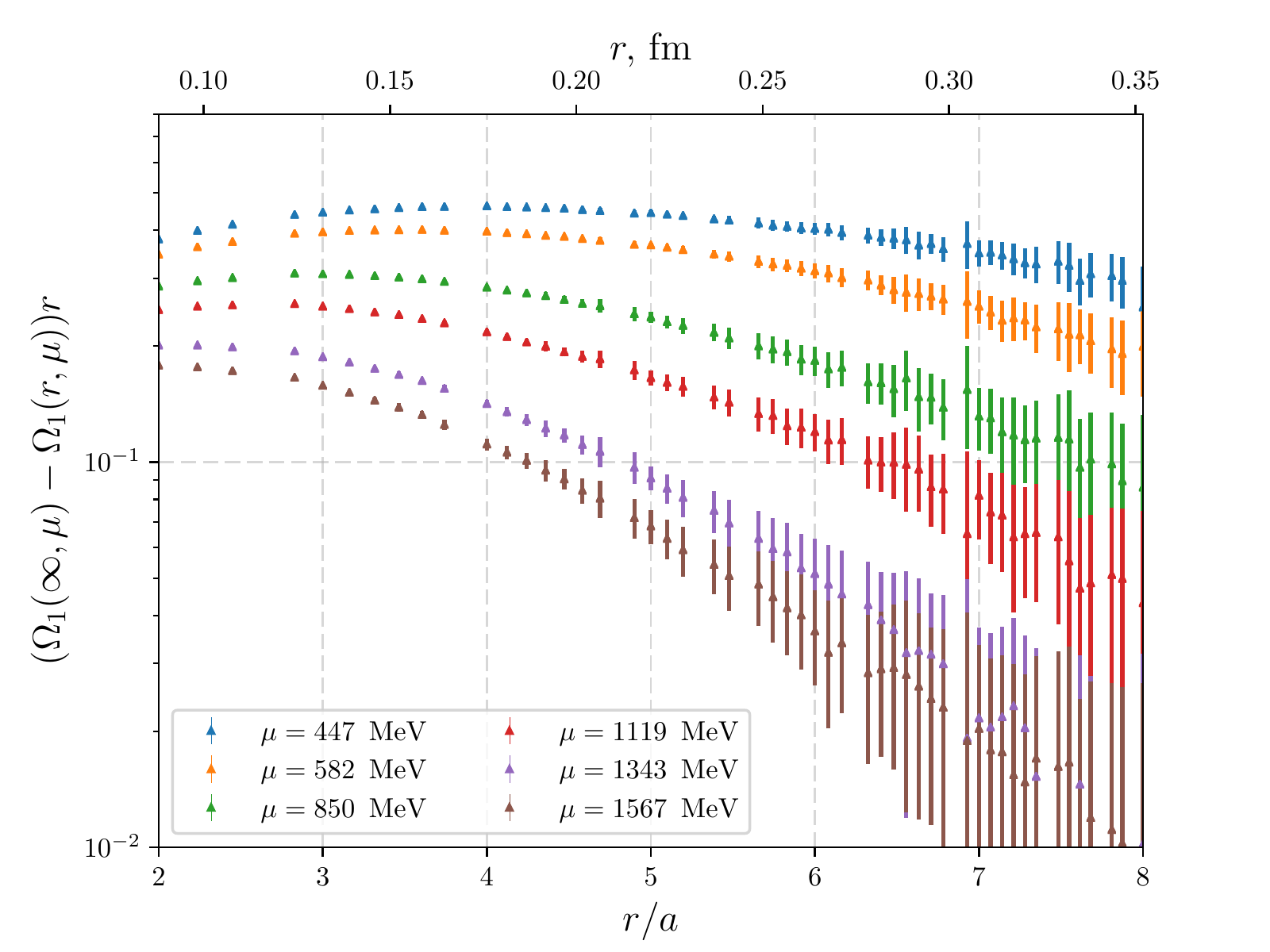}
	\caption{The expression $(\Omega_1(\infty, \mu) - \Omega_1(r, \mu)) r$ in logarithmic scale as a function of distance for various $\mu$.}
	\label{fig:omegalog}
\end{figure}

The scale of the Debye screening in perturbtion theory is denoted by the Debye mass, which to one-loop order (for the $N_c=2$) reads
\beq
m_D^2(\mu) = \frac {4} {\pi} \alpha_s(\mu) \mu^2\,.
\label{md}
\eeq
To describe or results for the $R_{sc}$ it is reasonable to assume that the screening length is inversely proportional to $m_D(\mu)$. 
For this reason we fit our data by the formula 
\beq
R_{sc}=\frac 1 {A m_D(\mu)}\,,
\label{rsc_debye}
\eeq
where the $A$ is some factor. We fit our data in the region $\mu \in (900, 1800)~$MeV and use a two-loop approximation for the running of the coupling constant 
$\alpha_s(\mu)$ (see formula (\ref{g_two_loops}) with $N_f=N_c=2$). The fit describes our data well ($\chi^2/dof\simeq 0.8$) and the best fit parameters are 
$A=1.4 \pm 0.4$, $\Lambda = 140 \pm 80~$MeV. In the region $\mu>1800~$MeV the data cannot be described by formula (\ref{rsc_debye}).

\begin{figure}[b]
\begin{minipage}[t]{0.48\textwidth}
	\centering
	\includegraphics[scale=0.5]{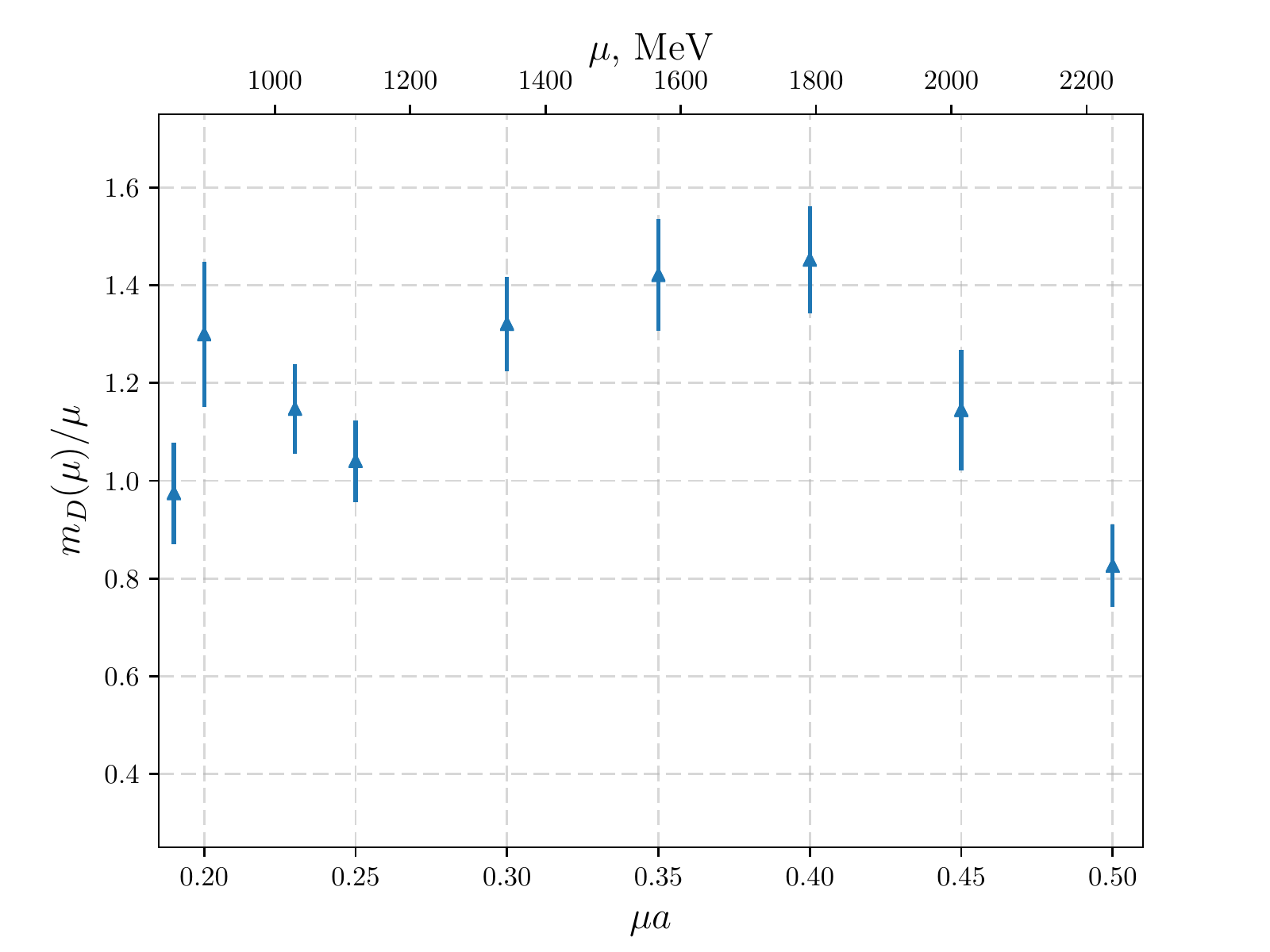}
	\caption{The ratio $m_D/\mu$ as a function of the chemical potential calculated from the fit of lattice data by formula (\ref{omega_large_d}).}
	\label{fig:mD}
\end{minipage}
	\hfill
\begin{minipage}[t]{0.48\textwidth}
	\centering
	\includegraphics[scale=0.5]{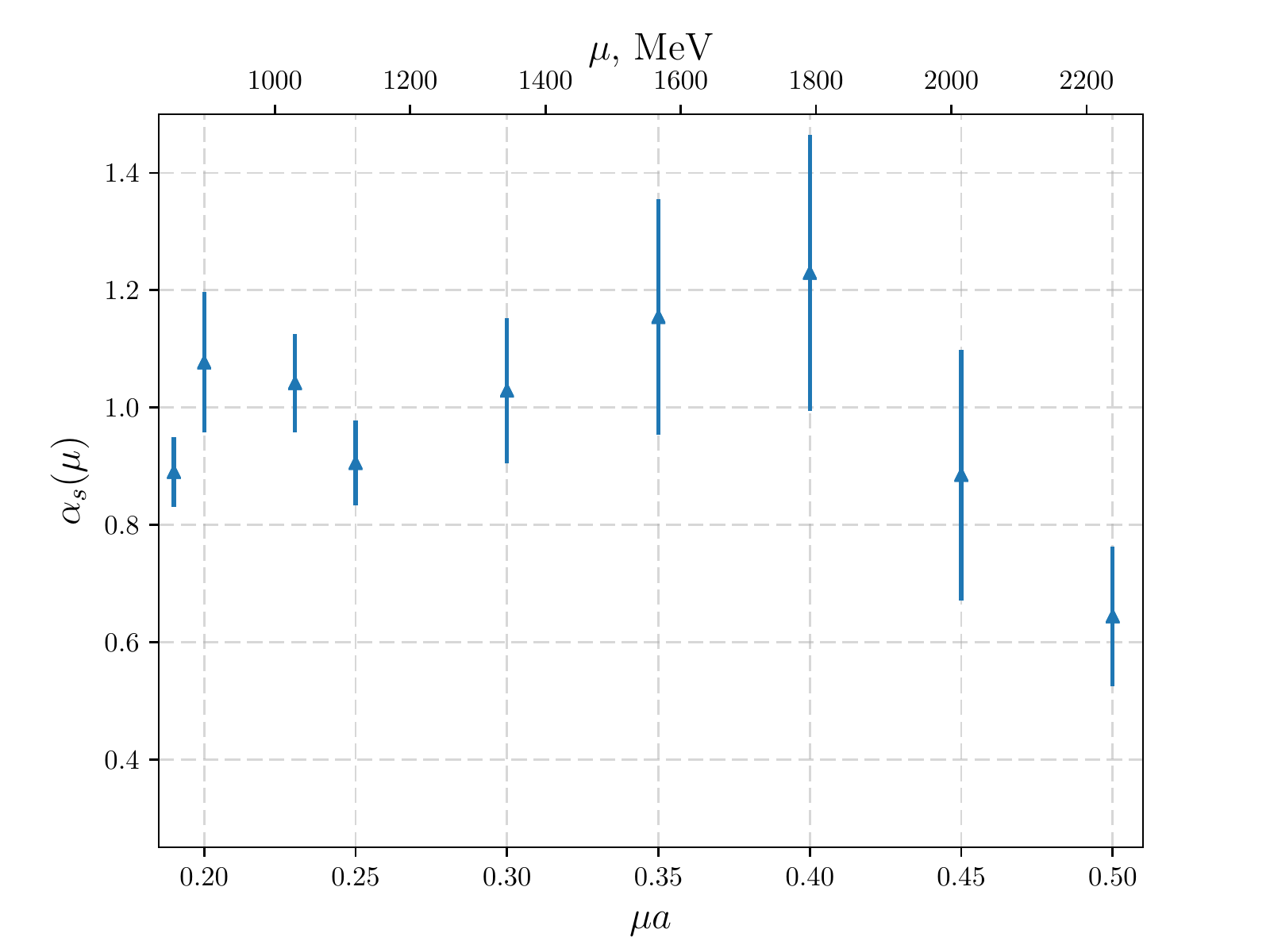}
	\caption{The strong coupling constant $\alpha_s$ as a function of the chemical potential calculated from the fit of lattice data by formula (\ref{omega_large_d}).}
	\label{fig:alphas}
\end{minipage}
\end{figure}

Now let us study how the Debye screening phenomenon manifests itself in the large distance behavior ($r\mu \gg 1$) of the grand potential.
In this case the dominant scale is the chemical potential, i.e. the running coupling constant depends only on  $\mu$: $g(r,\mu)=g(\mu)$. For sufficiently large density one can apply perturbation theory to calculate grand potentials. Perturbatively the grand potential $\Omega_{\bar q q}(r, \mu)$ 
is determined by two-gluon exchange and it is rapidly decreasing with distance function. Contrary to 
$\Omega_{\bar q q}(r, \mu)$ the color singlet grand potential $\Omega_1(r,\mu)$ is determined by one-gluon exchange. 
In this paper we consider only $\Omega_1(r,\mu)$, whose leading order contribution has the form
\beq
\Omega_1(r,\mu)=\Omega_1(\infty,\mu)-\frac 3 4 \frac {\alpha_s(\mu)} r e^{-m_D r}\,,
\label{omega_large_d}
\eeq
where $m_D$ is the Debye mass given by the expression (\ref{md}). It tells us that due to Debye screening at sufficiently large distance 
the expression $(\Omega_1(\infty, \mu) - \Omega_1(r, \mu)) r$ is an exponentially decreasing function of the distance. 
We plot $(\Omega_1(\infty, \mu) - \Omega_1(r, \mu)) r$ in logarithmic scale in figure~\ref{fig:omegalog}. From this 
figure the exponential decrease at large distance is seen starting from the $\mu=850~$MeV what confirms Debye screening phenomenon 
in deconfined dense quark matter. The deviation from a purely Coulombic Debye-like behavior at intermediate distances may be related to the remnants of the string, which is not perfectly screened.

Further we fit our data in the deconfinement phase for $\Omega_1(r, \mu)$ at sufficiently large $r$ by the formula (\ref{omega_large_d}). 
The results for $m_D/\mu$ and $\alpha_s(\mu)$ as a function of the chemical potential are shown in figure~\ref{fig:mD} and figure~\ref{fig:alphas}. 
From figure~\ref{fig:mD} it is seen that the dependence of the Debye mass on the chemical potential is $m_D \sim \mu$. Due to large uncertainties of the calculation, we are not able to resolve the running of the coupling constant with $\mu$. For the same 
reason we are not able to observe the running of the $\alpha_s$, as shown in figure~\ref{fig:alphas}. The running coupling 
is constant within the uncertainty of the calculation up to the $\mu<1800~$MeV and it starts to drop in the region $\mu>1800~$MeV. 

From figure~\ref{fig:alphas} one sees that the coupling constant is of order of unity $\alpha_s \sim 1$. 
It means that for all densities the system under study is strongly correlated. In addition one can 
expect that the one-loop formula  for the Debye mass (\ref{md}) is considerably modified by higher order radiative corrections. In the deconfined phase at finite temperature and zero density one also obtains a large coupling constant (see e.g. \cite{Kaczmarek:2004gv, Kaczmarek:2005ui}). 

\begin{figure}[h!]
	\centering
	\includegraphics[scale=0.65]{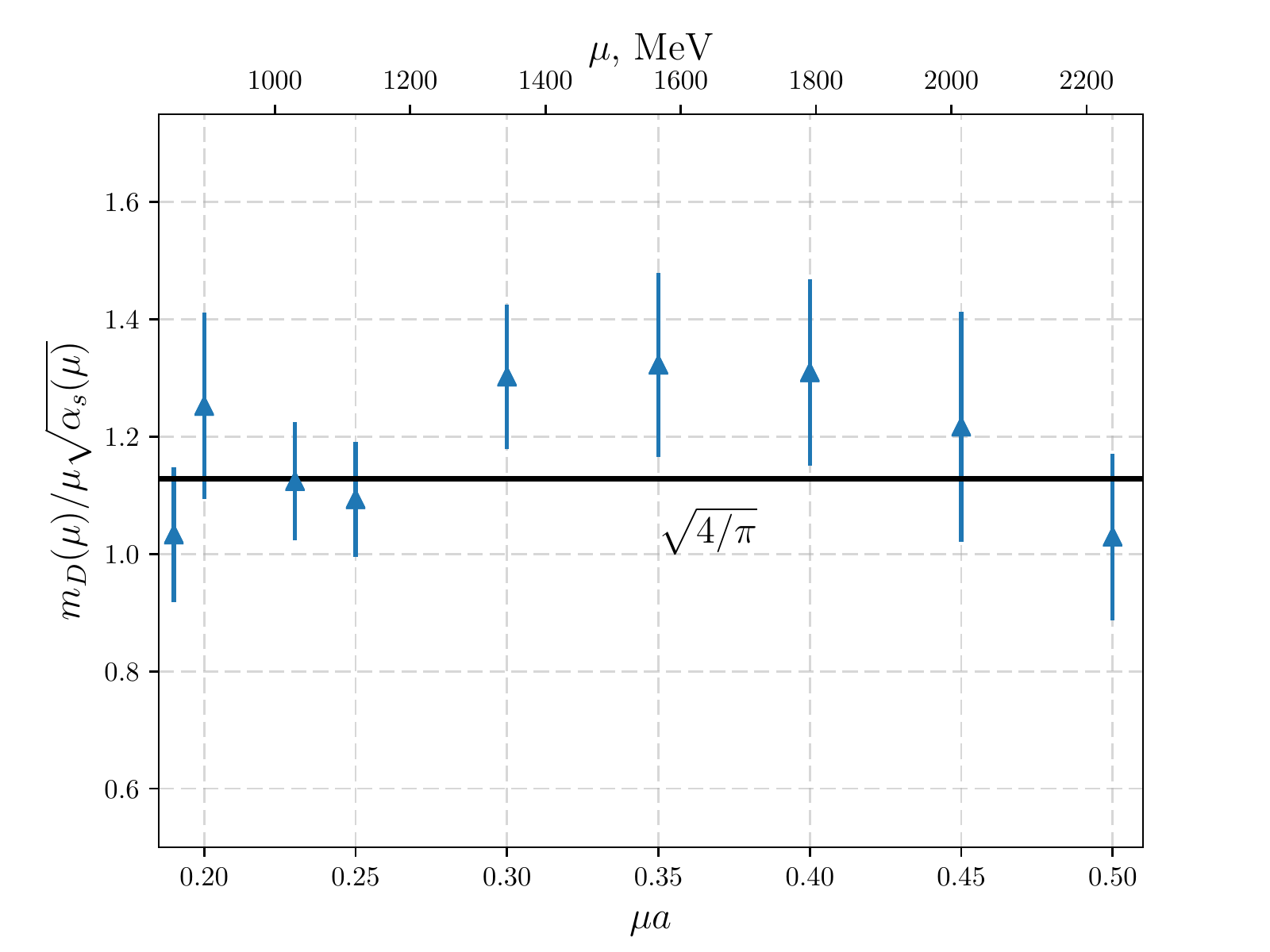}
	\caption{The expression $m_D(\mu) / (\mu \sqrt{\alpha_s(\mu)} )$ as a function of chemical potential. The $m_D(\mu)$ and the $\alpha_s(\mu)$ are extracted from fit of the $\Omega_1(r, \mu)$ data by formula (\ref{omega_large_d}). At the leading order approximation the ratio equals to the constant $\sqrt{4 / \pi}$ shown as a black horizontal line.}
	\label{fig:ratio}
\end{figure}

Despite the large coupling constant, it turns out that the one-loop formula (\ref{md}) works quite well. 
To show this, we compute the ratio $m_D/(\mu \sqrt {\alpha_s})$. At the leading order approximation this ratio is 
$\sqrt {4/\pi}$. In figure~\ref{fig:ratio} we plot this ratio and find that within the uncertainties of the calculation formula (\ref{md}) works quite well for $m_D$ and $\alpha_s$ extracted from the 
color singlet grand potential.

\section{Conclusion and discussion}
\label{sec:conslusion}

In this paper we continued our study of two-color QCD
at finite density and low temperature based on lattice simulations.
Our simulations were performed on 32$^4$ lattices with rooted staggered fermions at a relatively small lattice spacing a=0.044 fm, which allowed us to study two-color QCD very large baryon densities (up to quark chemical potential $\mu>2000~$MeV) while avoiding strong lattice artifacts.

The aim of the present paper was the study of the interaction between a static quark-antiquark pair in two-color dense quark matter. To this end we performed computations of the Polyakov loop correlation functions 
and calculated the color averaged, color singlet and color triplet grand potentials. 
To handle appropriately the divergent self-energy contribution to the Polyakov loop correlation functions, we conduct renormalization through a matching of the color singlet grand potential to the static interaction 
potential of quark-antiquark pair at short distances. Having determined the renormalized grand potentials, we calculated 
the renormalized grand potential of a single quark/antiquark and average Polyakov loop. 
In addition we calculated the quark number induced by a static quark antiquark pair and its internal energy.

The confinement/deconfinement transition at finite density manifests itself in an increasing value of the Polyakov loop. The finite density transition does not show a region of rapid rise of the Polyakov loop contrary to the finite temperature case. For this reason we conclude that the transition from confinement to deconfinement at finite density is smooth. 

In addition we calculated the screening length $R_{sc}$ which is defined as 
\beq
V_{\mu=0}(R_{sc})=\Omega(\infty, \mu),
\eeq
where $V_{\mu=0}(r)$ is the static potential at zero density and the $\Omega(\infty, \mu)$ is the 
grand potential of a static quark-antiquark pair at infinite distance. 
In the confined phase, the screening length is determined by the string breaking length, whereas
in the deconfined phase $R_{sc}$ is determined by the Debye screening phenomenon. 

The result of the calculation of the screening length shows that consistent with intuition, the larger the chemical potential the smaller $R_{sc}$, the string breaking distance.
We believe that the decrease of the string breaking distance with density can be attributed to a further string breaking mechanism in dense matter. In dense two-color QCD, in addition to the possibility 
to break the string by a quark-antiquark pair, it becomes possible to break the string by two quarks
which can be extracted from a two-color baryon -- the scalar diquark. Notice also that 
it does not cost any energy to remove the scalar diquark from the condensate and break the string.
As the result of this phenomenon, 
after the string breaking one end up with one heavy-light meson and one heavy-light diquark. 
Lattice studies show \cite{Braguta:2016cpw} that in the region $\mu>m_{\pi}/2$ the scalar diquark condensate increases with the chemical potential, i.e. it becomes easier to find two quarks and to break the string. 

If one increases the chemical potential then at some density $R_{sc}$ becomes so small 
that the string cannot be created at all. Once created it will be immediately broken by the 
two-color baryons -- the scalar diquarks. This is our hypothesis of the deconfinement mechanism in two-color dense quark matter.

The behavior of the string breaking distance in dense matter and the deconfinement mechanism 
are not specific only for two-color QCD. We believe that a similar process can be realized in SU(3) QCD with the difference that one has to replace two-quark baryon in SU(2) by three-quark baryon in SU(3). 
In particular, one can expect that the screening length which has the same definition as in two-color 
QCD is decreasing function of the chemical potential. in turn, the larger the density the smaller the string breaking distance. For three colors this behavior can be explained as follows: at nonzero chemical potential one has a nonzero baryon density in the system. Baryons 
which form this density can break the string, splitting it into one quark and one diquark. 
Notice that one does not need additional energy to create the baryon since the baryons are already present, due to the nonzero chemical potential. After the string breaking 
one has one heavy-light meson and heavy-light baryon. Finally the larger the 
the chemical potential, the larger the number of baryons which can break the string,
i.e. the string breaking distance is a decreasing function of the chemical potential. 

Notice that in three-color QCD one might also have a similar mechanism of deconfinement 
at finite density, as we proposed above for two-color. In particular, deconfinement 
takes place at the density at which the $R_{sc}$ is so small that the string 
cannot be created.  

In the previous section we considered the large distance behavior of the color singlet grand 
potential in the deconfined phase. In analogy with Debye screening in the Abelian theory and using leading order perturbation theory, we attempt to quantitatively describe the observed behavior and find good agreement with the lattice data.

We calculated the Debye mass and the coupling 
constant for various chemical potentials. The coupling constant extracted in this way takes on values $\alpha_{s} \sim 1$, which tells us that despite the large baryon density, the system remains strongly coupled. It was also found that despite the large coupling 
constant the one-loop formula for the Debye mass works well at large distances within the uncertainty 
of the calculation. 

In this paper we found that the region $\mu<2000~$MeV physically differs from the region $\mu>2000~$MeV. 
This manifests itself in different behavior of the following observables: the Polyakov line, the grand potential, the screening length $R_{sc}$, the Debye mass and effective coupling constant. 

While we do not yet fully understand the physics, which is responsible for this behavior, one possibility is that the value of the chemical potential $\mu \sim 2000~$MeV is exceptional since it divides the region $\mu<2000~$MeV with 
a spatial string tension from that at $\mu > 2000~$MeV where it vanishes. This may imply that the point $\mu \sim 2000~$MeV separates 
systems with and without magnetic screening. Further study in this direction is required. 

Finally we are going to discuss lattice artifacts which result from the saturation effect. 
It is known that at large values of the chemical potential $a\mu \sim 1$ a saturation effect starts 
to be seen. The essence of this effect is
that all lattice sites are filled with fermionic degrees of freedom, and it is not possible to put
more fermions on the lattice (``Pauli blocking''). It is known that the saturation effect is
accompanied by the decoupling of the gluons from fermions. Thus, effectively due to saturation, 
the system becomes simply gluodynamics, which is confined at low temperatures. From this
consideration it is clear that in order to study the properties of quark matter at 
large baryon density one should have sufficiently small lattice spacing such that the properties 
are not spoiled by this kind of artificial confinement at large values of the chemical potential. 
We believe that because of the saturation effect the deconfinement in dense SU(2)
matter has not been seen before.

The results of the study presented in this paper are obtained for the chemical potentials $\mu < 2200~$MeV ($a\mu \leq 0.5)$.
We believe that our results are not spoiled by saturation for the following reasons. 
First, for the $\mu > 2000~$MeV (up to $\mu \sim 2500~$ MeV~\cite{Bornyakov:2017txe}) 
the spatial string tension is vanishing. Second, we do not see a respective rise 
of the timelike string tension. 
Moreover, the static potential for $\mu > 2000~$MeV (up to $\mu \sim 2500~$ MeV~\cite{Bornyakov:2017txe}) 
is well described by Debye screening potential. 
So, the properties of the system in the range $\mu > 2000~$MeV are
very different from those of plain gluodynamics at small temperatures. 

Notice also that in our previous study of dense two-color QCD~\cite{Braguta:2016cpw} we found that
the onset of the saturation effects are seen at $a\mu \sim 0.7-0.8$. This was deduced through the decrease of the diquark
condensate for $a\mu>0.7$ (while it is rising
with $\mu$ for the $a\mu < 0.7$). The rise of the diquark condensate in
the continuum is related to the rise of the
Fermi surface. The decrease of the diquark condensate on the lattice
is evidently related to the onset of the saturation effect what can be seen as follows.
Due to finite number of the fermion states in the lattice Brillouin
zone there is a chemical potential from which
the rise of the chemical potential does not lead to the rise of the
Fermi surface on the lattice. Notice that at this value of the chemical potential 
not all fermion states on the lattice are filled and the saturation takes place at larger 
values of the chemical potential. 

Finally the deviation of the lattice measured baryon density from the
baryon density calculated for free fermions
is 10 \% for $a\mu=0.45$(~2000~MeV) and 20\% for
$a\mu=0.50$(~2250 MeV). We argue that such a deviation
 even if it could be attributed to saturation cannot lead
to considerable modification of physics. Notice also that such a deviation
may also be explained by other mechanism than saturation, e.g. the
finite lattice spacing which is present for any $a\mu$.

Taking into account all what is written above we believe that in the region under consideration in this paper, ($a\mu<0.5$)
our results are not spoiled by eventual saturation effects. Notice that the strict proof of last statement requires additional lattice simulations at smaller lattice spacing, which are planned in the future.

\appendix

\section{The interaction potential of  static quark-antiquark pair at zero density}
\label{appA}
In Appendix A we are going to calculate the interaction potential 
of static quark-antiquark pair $V(r)$ at zero density through lattice measurement of Wilson loops. 
In this paper we use the $V(r)$ in order to renormalize the correlation functions of the Polyakov lines. 

For the calculation of Wilson loops we have employed one step of HYP 
smearing~\cite{Hasenfratz:2001hp} for temporal links with the smearing 
parameters according to the HYP2 parameter set~\cite{DellaMorte:2005nwx},  
followed by 100 steps of APE smearing~\cite{Albanese:1987ds} for spatial 
links only with the smearing parameter $\alpha_{APE} = 2/3$. The similar 
smearing scheme was applied in the paper~\cite{Bonati:2014ksa} for the 
extraction of $V(r)$ from the Wilson loops. In the case of spatial 
Wilson loops (see below) the smearing scheme was adopted respectively to 
consider one of the spatial directions as a ``temporal direction''
\footnote{ The calculation of the other gluon observables, like the correlation function 
of Polyakov loops, color singlet/triplet free energy and etc., one 
step of HYP smearing with the same parameters was employed. }.

Having measured Wilson loops for all  distances between static charges
one can calculate the interaction potential $V(r)$. Notice, however, 
that the interaction potential is determined up to a renormalization constant.
We find this renormalization constant for the interaction potential 
at $\mu=0$ using the following procedure.  
It is known that for the distances $r>0.5 \mathrm{~fm}$~\cite{Sommer:1993ce} the interaction
potential is well described by the linear confinement potential corrected by  
the so-called L\"uscher term which describes string fluctuations
\beq
V(r)=\sigma r - \frac {\pi} {12 r}.
\label{potential}
\eeq
In order to calculate the interaction potential unambiguously
we fit our lattice data by the potential $V(r)=\sigma r - \pi/ 12r + C$.
The fit is good and the constant is equal to $C=726\pm 13~\mbox{MeV}$.
The renormalization constant $C$ determines the energy shift of 
the lattice potential as compared to (\ref{potential}). 
The renormalized lattice potential can be obtained 
through the shifting of the lattice potential down by $C$ and it reproduces 
the potential (\ref{potential}) for $r>0.5 \mathrm{~fm}$.  

At the end of this section we would like to mention that the potential $V(r)$ calculated as it was 
described above contains lattice artifacts at small distances $r/a \leq 3$. These artifacts result from 
the HYP smearing which modifies the potential at small distances and the violation of the rotation 
invariance by our lattice at short distances. However, the measurements of the correlators of the 
Polyakov lines and the Wilson loops are carried out with the same HYP smearing and on the same 
lattice. For this reason, the interaction potential determined in this section is 
appropriate for the renormalization of the Polyakov lines correlators. 

\section{Quarkonia properties in two-color QCD}
\label{appB}
In Appendix B we are going to estimate the two-color quarkonia masses and their sizes using potential model.  
The background of all potential models is the interaction potential. In this paper 
we are going to use the Cornell potential~\cite{PhysRevD.17.3090} of the form
\beq
V(r) = - \frac{3}{4}\frac{\alpha_s}{r} + \sigma r.
\label{Cornell}
\eeq
Notice that the coefficient $3/4$ is front of the Coulomb term is due to $N_f=2$ colors in our system. 
In the calculation we use the string tension $\sigma = (476\,\mbox{MeV})^2$ which was calculated in this paper. 
The effective coupling constant $\alpha_s$ is extracted from the Cornell potential fit of our lattice data in the 
region $3 \leqslant r/a \leqslant 12$. Thus we obtained the following value $\alpha_s = 0.31$. 
Notice that this value a larger that that in SU(3) theory $\alpha_s=0.21$~\cite{Kaczmarek:2005ui}. 
We believe that the difference between two- and three-color QCD can be attributed to different RG running of the coupling constants. 
In particular, the two loops running of the coupling constant is given by the formula
\beq
g^{-2}(\mu) = 2 \beta_0 \log\frac{\mu}{\Lambda} + \frac{\beta_1}{\beta_0}\log\left( 2 \log\frac{\mu}{\Lambda} \right),
\label{g_two_loops}
\eeq
with $\beta_0 = (4 \pi)^{-2} \left(\frac{11}{3} N_c - \frac{2}{3} N_f\right)$, $\beta_1 = (4 \pi)^{-4} \left(\frac{34}{3} N_c^2 - \frac{N_c^2 - 1}{N_c} N_f - \frac{10}{3} N_c N_f \right).$ 
The running is controlled by the coefficients $\beta_0, \beta_1$ which are different for $N_c=2$ and $N_c=3$. 
More quantitatively if one takes $\Lambda = 200\,\mbox{MeV}$, $\mu = 2\,\mbox{GeV}$, for the $N_f = 2$ and $N_c = 2$ one has $\alpha_s = 0.35$, whereas for the $N_c = 3$  $\alpha_s = 0.22$.

We use the non-relativistic Schr\"odinger equation with the potential~\ref{Cornell} in order to estimate the quarkonia masses and sizes. In the calculation we take the charm quark mass $m_c = 1850\,\mbox{MeV}$ from the paper~\cite{PhysRevD.17.3090}. Having solved the equation numerically, one obtains the following values of the states energies and mean squared distance between quarks $\sqrt{\langle r^2 \rangle}$:
\begin{table}[h]
\centering
    \begin{tabular}{ | c | c | c | c |}
    \hline
    $ $ & $J/\psi$ & $\chi_c$ & $\psi'$ \\ \hline
    E, MeV & 551.56 & 925.60 & 1132.60 \\ \hline 
    $\sqrt{\langle r^2\rangle}$, fm & 0.41 & 0.61 & 0.76 \\ \hline
    \end{tabular}
\end{table}

\acknowledgments

V.\,V.\,B. acknowledges the support from the BASIS foundation. N.\,Yu.\,A. acknowledges the support from the BASIS foundation and FAIR-Russia Research Center. The work of A.\,Yu.\,K. was supported by FAIR-Russia Research Center and RFBR grant 18-32-00071. The work of A.\,A.\,N. was supported by RFBR grant 18-32-00104. V.\,G.\,B. acknowledges the support by RFBR grant 16-02-01146. A.\,R. acknowledges partial support by the DFG funded collaborative research center SFB1225 ``ISOQUANT''. This work has been carried out using computing resources of the federal collective usage center Complex for Simulation and Data Processing for Mega-science Facilities at NRC ``Kurchatov Institute'',~\url{http://ckp.nrcki.ru/}. In addition, the authors used the equipment of the shared research facilities of HPC computing resources at Lomonosov Moscow State University, the cluster of the Institute for Theoretical and Experimental Physics and the supercomputer of Joint Institute for Nuclear Research ``Govorun''.

\bibliographystyle{JHEP}
\bibliography{bibliography}

\end{document}